\documentstyle[12pt,epsfig]{article}
\topmargin - 1.0 true cm
\newcommand{\beq}{\begin{equation}}
\newcommand{\eeq}{\end{equation}}
\def\bea{\begin{eqnarray}}
\def\eea{\end{eqnarray}}
\def\nn{\nonumber}
\def\barr{\begin{eqnarray}}
\def\earr{\end{eqnarray}}

\def\lsim{\raise0.3ex\hbox{$\;<$\kern-0.75em\raise-1.1ex\hbox{$\sim\;$}}}
\def\gsim{\raise0.3ex\hbox{$\;>$\kern-0.75em\raise-1.1ex\hbox{$\sim\;$}}}

\def\dm{\Delta m}

\def\gh{\Gamma_H}
\def\gl{\Gamma_L}
\def\dg{\Delta \Gamma_d}
\def\aa{{\cal A}}
\def\lt{\left}
\def\rt{\right}
\newcommand{\adi}{{\cal A}_{\rm CP}^{\rm dir}}
\newcommand{\ami}{{\cal A}_{\rm CP}^{\rm mix}}
\newcommand{\adg}{{\cal A}_{\rm \Delta\Gamma}}

\begin{document}
\baselineskip=6truemm

\begin{flushright}
hep-ph/0109088 \\
CERN-TH/2001-200\\
IFP-796-UNC\\
MPI-PhT/2001-27
\end{flushright}

\begin{center}
\bigskip

{\Large {\bf 
Measurement of the Lifetime Difference of $B_d$ Mesons:\\ 
\vspace{0.5cm}
Possible and Worthwhile?}}

\vspace{1cm}

\end{center}
 
\bigskip\bigskip

\begin{center}
{\large \bf 
A. S. Dighe$^{a,}$\footnote{amol.dighe@cern.ch},   
T. Hurth$^{b,}$\footnote{tobias.hurth@cern.ch},
C. S. Kim~$^{c,}$\footnote{cskim@mail.yonsei.ac.kr},~~ 
and~~ 
T. Yoshikawa~$^{d,}$\footnote{tadashi@physics.unc.edu}}
\end{center}

\begin{center}

{
$a$: {\it Max-Planck-Institut f\"ur Physik, F\"ohringer Ring 6,
        D-80805 M\"unchen, Germany}

$b$: {\it Theory Division, CERN, CH-1211 Geneva 23, Switzerland}\\

$c$: {\it Department of Physics and IPAP, Yonsei University, 
Seoul 120-749, Korea}\\

$d$: {\it Department of Physics and Astronomy,
          University of North Carolina, \\
        Chapel Hill, NC 27599-3255, USA }
}

\end{center}

\centerline{(\today )}

\begin{abstract}

\noindent
We estimate the decay width difference $\dg/\Gamma_d$ in the 
$B_d$ system including $1/m_b$ contributions and part
of the next-to-leading order QCD corrections, 
and find it to be around  $0.3\%$.
We explicitly show that the time measurements of an 
untagged $B_d$ decaying to a single final state isotropically 
can only be sensitive to quadratic terms in
$\dg/\Gamma_d$, and hence the use of at least two different
final states is desired. 
We discuss such pairs of candidate decay channels 
for the final states and explore the feasibility of a
$\dg/\Gamma_d$ measurement through them. 
The measurement of $\dg$ would be essential for an accurate 
measurement of $\sin(2\beta)$ at the LHC. 
The nonzero width difference 
can also be used to identify new physics effects and 
to resolve a twofold discrete ambiguity in the
$B_d$--$\bar{B}_d$ mixing phase.
We also derive an upper bound on the value of $\dg/\Gamma_d$
in the presence of new physics, and point out some
differences in the phenomenology of width differences in the 
$B_s$ and $B_d$ systems.
\end{abstract}

\newpage

\section{Introduction}

Within the standard model (SM), the difference in the decay 
widths of $B_d$ mesons is CKM-suppressed with respect to 
that in the $B_s$ system. A rough estimate leads to
\beq
\frac{\Delta \Gamma_d}{\Gamma_d} \sim
\frac{\Delta \Gamma_s}{\Gamma_s} \cdot \lambda^2
\approx  0.5 \% ~~,
\label{dg-approx}
\eeq
where $\lambda = 0.225$ is the sine of Cabibbo angle, and we have taken 
$\Delta \Gamma_s/\Gamma_s \approx 15\%$ \cite{aleksan,BBD,BBGLN}. 
Here $\Gamma_{d(s)} = (\Gamma_L + \Gamma_H)/2$ is the average decay width 
of the  light and heavy $B_{d(s)}$ mesons ($B_L$ and $B_H$ respectively).
We denote these decay widths by $\Gamma_L,
\Gamma_H$ respectively, and define $\Delta \Gamma_{d(s)} 
\equiv \gl - \gh$. 
No experimental measurement of $\Delta \Gamma_d$ is
currently available. Moreover, no motivation for
its measurement (other than just measuring another number to
check against the SM prediction) has been discussed, and hence  
the study of the lifetime difference between $B_d$ mesons 
has hitherto been neglected as compared to that in the
$B_s$ system. The phenomenology of the lifetime difference 
between $B_s$ mesons has been explored in detail in
\cite{isi95,dfn}.

With the possibility of experiments with high time
resolution and high statistics, it is worthwhile to have a look
at this quantity and make a realistic estimate of the 
possibility of its measurement.
At LHCb for example, the proper time resolution is expected to be
as good as $\Delta \tau \approx 0.03$ ps.
This indeed is a very small fraction of the $B_d$ lifetime
($\tau_{B_d} \approx 1.5$ ps \cite{pdg}), 
so the time resolution
is not a limiting factor in the accuracy of the measurement,
the statistical error plays the dominant role.
Taking into account the estimated
number of $B_d$ produced --- for example the number of 
reconstructed $B_d \rightarrow J/\psi \, K_S$ events at the LHC is expected
to be $5 \times 10^5$ \, (\cite{lhc} table 3) ---  the measurement of the
lifetime difference does not look too hard at first glance.
Naively, one may infer that if the number of relevant events
with the proper time of decay measured with the
precision $\Delta \tau$ is $N$, then the value
of $\Delta \Gamma_d / \Gamma_d$ is measured with an
accuracy of $1 / \sqrt{N}$. With a
sufficiently large number of events $N$, it should be
possible to reach the accuracy of 0.5\%
or better.

The measurement of $\dg / \Gamma_d$ is in reality harder
than what the above naive expectation may suggest,
since most of the quantities that involve the lifetime
difference are quadratic in the small quantity $\dg / \Gamma_d$.
In fact, as we shall explicitly show in this paper, 
the time measurements in the decays of an untagged $B_d$ to a single
final state are sensitive only to $(\dg/\Gamma_d)^2$.
This implies that
in order to discern two different lifetimes, the measurements
need to have an accuracy of 
$(\dg/\Gamma_d)^2 \sim 2.5 \times 10^{-5}$,
which is beyond the reach of the currently planned experiments.

However, 
the combination of lifetimes measured in two different untagged decay 
channels may be sensitive to linear terms in $\dg/\Gamma_d$. 
We explore three
pairs of such untagged measurements in this paper: (i) lifetime measurements
through decays to self-tagging (e.g. semileptonic)
final states and to CP eigenstates,
(ii) CP even and odd components in the decay mode 
$B_d \to J/\psi K^*(K_s \pi^0)$, and
(iii) time-dependent untagged asymmetry between $B_d \to J/\psi K_S$
and  $B_d \to J/\psi K_L$.

The conventional ``gold-plated'' decays for 
$\beta$ measurement, $J/\psi K_S$ and $J/\psi K_L$, 
neglect the lifetime difference while determining 
$\sin(2 \beta)$. For an accurate determination of $\beta$, the
systematic errors due to $\Delta\Gamma_d/\Gamma_d$ need to be taken
into account. 
Moreover, there is the possibility that 
the measurement of the lifetime difference leads to a clear
signal for new physics. Furthermore, 
if the lifetime difference is neglected,
 the ambiguity 
$\beta \leftrightarrow (\pi/2 - \beta)$ remains unresolved. 
Observables that are sensitive to the
lifetime difference may resolve this discrete ambiguity
under certain conditions.

The observables mentioned above can also
give an independent measurement of $\cos(2\beta)$ in principle.
In order to be able to do this, however,  
the theoretical uncertainties on $\Delta\Gamma_d$
need to be minimized. 
Therefore, we start by presenting in Sec.~\ref{nll} 
a detailed calculation of $\dg$, including $1/m_b$ contributions
and part of the next-to-leading order (NLO) QCD corrections.
The NLO precision in the width difference 
$\dg$ is also essential for obtaining a proper matching of
the Wilson coefficients to the matrix elements of local operators
from the lattice gauge theory.

The rest of the paper is organized as follows.
In Sec.~\ref{phenom} we explicitly demonstrate
the quadratic dependence on $\dg/\Gamma_d$
of quantities measurable through untagged $B$ decays to a 
single final state.
We explore the combinations of decay modes that can measure
quantities linear in $\dg/\Gamma_d$.
We calculate the corrections due to $\dg$ 
to the measurement of $\sin(2\beta)$ through $B_d \to J/\psi K_S$,
and also indicate the possibility of the $\dg$ measurement 
through tagged decays to CP eigenstates. 
In Sec.~\ref{contrast}, we point out important 
differences in the upper bounds on $\Delta\Gamma_s$ and $\dg$ in
the presence of new physics, and elaborate on the possibility
of detecting new physics and resolving discrete ambiguities 
in the mixing phases through them.
We summarize our findings in Sec.~\ref{concl}.

\section{Estimation of $\dg$}
\label{nll}

\subsection{Basic definitions}

We briefly recall the basic definitions:
in the Wigner--Weisskopf approximation 
the oscillation and the decay of a general linear combination
of the neutral flavour eigenstates $B_d$ and $\bar{B}_d$,\,
$a |{B_d}\rangle  + b |{\bar B_d}\rangle $,\,
is described  by the time-dependent Schr\"odinger equation
\begin{equation}
i{d\over dt}\pmatrix{a\cr b\cr}=
\left({\bf M}-{i\over2}{\bf \Gamma}\right)\pmatrix{a\cr b\cr}.
\end{equation}
Here ${\bf M}$ and ${\bf \Gamma}$ are $2\times2$ Hermitean matrices.
CPT invariance leads to the conditions ${ M_{11}} = { M_{22}}$
and  ${ \Gamma_{11}} = { \Gamma_{22}}$.
Exact CP invariance would imply ${ M_{21}} = { M_{12}}$
and ${ \Gamma_{21}} = { \Gamma_{12}}$ (a phase 
choice, namely ${\cal CP}|{B_d}\rangle = - |{\bar B_d}\rangle $$,
{\cal CP}|{\bar B_d}\rangle = - |{B_d}\rangle $ is made). 
Independent of the choice of the unphysical phases, CP invariance
(in mixing) would imply ${\rm Im} ({ M^*_{21} \Gamma_{21}}) = 0$.

The mass eigenstates, the light $B_L$ and the heavy $B_H$, are 
given by 
\begin{equation}
|{B_{L,H}}\rangle =p|{B_d}\rangle \pm q|{\bar B_d}\rangle 
\label{p-q}
\end{equation}
with the normalization condition $|q|^2+|p|^2=1$. 
Only the 
magnitude $|q/p|$ is measurable, the phase of this quantity is
unphysical and can be fixed arbitrarily by convention.

The mass difference and the width difference between the 
physical states are defined  by
\begin{equation}
\Delta m  =    M_H-M_L, \quad \Delta\Gamma =
\Gamma_L - \Gamma_H ~,
\end{equation}
such that $\Delta m > 0, \dg >0$ in the SM.
The real and imaginary parts of the eigenvalue equations are the following:
\bea
(\Delta m)^2-{1\over4}(\Delta\Gamma)^2\ &=&\ (4|M_{21}|^2-|\Gamma_{21}|^2),\\
\Delta m\Delta\Gamma\ &=&\ - 4 {\rm Re} (M_{21}^* \Gamma_{21}).
\eea
With the help of  the  CP-violating parameter $\delta$     
\begin{equation}
\delta \equiv \frac{-2~ {\rm Im} ({ M^*_{21} \Gamma_{21}})}{(\Delta m)^2 + |\Gamma_{21}|^2} = |p|^2 - |q|^2 = \langle  B_L | B_H \rangle ,
\label{delta-def}
\end{equation}
The effect of CP violation due to  mixing 
on the mass difference $\Delta m$ and on the lifetime difference 
$\Delta \Gamma$ may be explicitly shown:
\bea 
\label{CPrep}
(\Delta m)^2 &=& \frac{4 |M_{21}|^2 - \delta^2 |\Gamma_{21}|^2}{1 + \delta^2}\\
(\Delta \Gamma)^2 &=& \frac{4 |\Gamma_{21}|^2 - 16 \delta^2 |M_{21}|^2}{1 + \delta^2}~.
\eea
In the limit of exact CP invariance ($\delta=0$)
the mass eigenstates coincide with the CP eigenstates, 
\, ${\cal CP} |B_H\rangle  = -  |B_H\rangle $
and ${\cal CP} |B_L\rangle  =  + |B_L\rangle $ and 
the mass difference and width difference are given by
$\Delta m = 2 |M_{21}|, \Delta \Gamma = 2 |\Gamma_{21}|$.
However, even with a non-zero $\delta$, 
taking into account that $\delta$ is constrained by the
upper bound $|\delta| \leq |\Gamma_{21}|/(2 |M_{21}|)$
and $\Gamma_{21}/M_{21} = {\cal O}(m_b^2/m_t^2)$, we can write
\begin{equation}
\label{MMrep}
\Delta m = 2 |M_{21}| \left[1 + {\cal O}
\left(\frac{m_b^4}{m_t^4}\right) \right], \quad
\Delta \Gamma = - \frac{ 2 {\rm Re} (M_{21}^* \Gamma_{21})}{|M_{21}|} 
\left[ 1 + {\cal O}\left(\frac{m_b^4}{m_t^4}\right) \right].
\label{Dapprox}
\end{equation}
We shall neglect the terms of ${\cal O}(m_b^4/m_t^4) \sim 10^{-6}$ 
in our calculations.

\subsection{Method of calculation}
\label{method}

In the following we consider the two off-diagonal elements 
$M_{21}$ and $\Gamma_{21}$,
which correspond respectively to  the dispersive and
the absorptive part of the transition amplitude from 
$B_d$ to $\bar B_d$. We follow the method of \cite{BBD,BBGLN}
which was used there in the $B_s$--$\bar B_s$ system 
(see also \cite{SM,SM2}).

\begin{figure}
\begin{center}
\epsfig{file=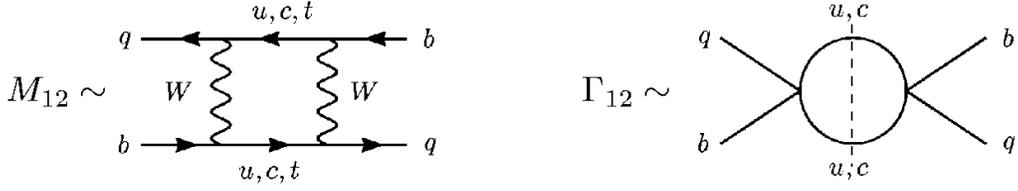,height=1in}
\end{center}
\caption{Schematic representation of Feynman diagrams 
for $M_{12} = M_{21}^*$ and $\Gamma_{12} = \Gamma_{21}^*$.
\label{feyn}}
\end{figure}

Within the SM the well-known box diagram is the starting point 
of the calculations. $M_{21}$ is related to the real part of this diagram
(see Fig.~\ref{feyn}).
The important QCD corrections  are most easily implemented with the help
of the standard operator product  expansion.
Because of the dominance  of the top quark contribution, 
$M_{21}$ can be  described by a local $\Delta B=2$ Hamiltonian
below the $m_W$ scale:
\barr
M_{21} & = & \frac{1}{2 M_{B_d}} \langle  \bar{B}_d | 
{\cal H}_{eff}^{\Delta B = 2} | B_d \rangle  \, 
 \left[ 1 + O\left(\frac{m_b^2}{m_W^2}\right) \right]~~,
\label{m12} \\
{\cal H}_{eff}^{\Delta B = 2} & =& 
 \frac{G_F^2}{16 \pi^2} (V_{tb}^* V_{td})^2   
C^Q (m_t,m_W,\mu) Q(\mu) + {\rm H.c.} ~~,\\
Q & = &(\overline{b}_i d_i)_{V-A} (\bar{b}_j d_j)_{V-A} ~~. 
\earr
The Wilson coefficient $C^Q$  contains the short-distance physics. It is
known up to NLO precision \cite{NLLM12}. The hadronic matrix element 
$\langle  \bar{B}_d | Q(\mu \approx m_b) | B_d \rangle $ 
will be discussed below.

In the standard model, $\Gamma_{21}$ is related to the imaginary part of the 
box diagram. Via the optical theorem it is fixed by the real 
intermediate states. Therefore, only the box diagrams with internal
$c$ and $u$ quarks contribute (see Fig.~\ref{feyn}).  
In contrast to the $B_s$--$\bar B_s$
case where the intermediate $c\bar c$ contribution is the dominating 
one, because of its CKM factor $(V_{cb}^* V_{cs})^2$, over 
the $u\bar u$, the $c\bar u$ and the $u\bar c$ contribution
(see Sec.~\ref{upper-bd}),
in the  $B_d$--$\bar B_d$  all four contributions have to be taken 
into account.
In the effective theory where we integrate out the $W$ boson, 
$\Gamma_{21}$ is given by: 
\bea
\Gamma_{21} = \frac{1}{2 M_{B_d}} 
\langle \bar{B}_d | {\rm Im} ~ i \int d^4x ~T ~{\cal H}^{\Delta B =1}_{eff}(x)
\, ~{\cal H}^{\Delta B =1}_{eff}(0) 
| B_d \rangle  ~~,
\eea
where
\bea
{\cal H}^{\Delta B =1}_{eff} &=& \frac{G_F}{\sqrt{2}} ( 
            V_{ub}^*V_{ud} \sum_{i=1,2}C_i Q_i^{uu} + 
            V_{cb}^*V_{ud} \sum_{i=1,2}C_i Q_i^{cu} +
            V_{ub}^*V_{cd} \sum_{i=1,2}C_i Q_i^{uc} + \nn\\ 
          &+& V_{cb}^*V_{cd} \sum_{i=1,2}C_i Q_i^{cc} 
          - V_{tb}^*V_{td} \sum_{i=3}^6 C_i Q_i ).
\label{HHH}
\eea
The operators are ($i,j$ denote color indices)
\bea
Q_1^{qq'} &=& (\overline{b}_iq_j)_{V-A} (\bar{q}'_jd_i)_{V-A},\,\,\,\,\,\, 
Q_2^{qq'} = (\overline{b}_iq_i)_{V-A} (\bar{q}'_jd_j)_{V-A},\\
Q_3 &=& (\overline{b}_id_i)_{V-A} (\bar{q}_jq_j)_{V-A},\,\,\,\,\,\,\,\,\,\, 
Q_4 = (\overline{b}_id_j)_{V-A} (\bar{q}_jq_i)_{V-A},\\
Q_5 &=& (\overline{b}_id_i)_{V-A} (\bar{q}_jq_j)_{V+A},\,\,\,\,\,\,\,\,\,\,   
Q_6 = (\overline{b}_id_j)_{V-A} (\bar{q}_jq_i)_{V+A}.
\eea
The penguin operators $Q_3$ -- $Q_6$  have small Wilson coefficients 
and are therefore suppressed with respect to the four-quark
operators -- which all have the same two Wilson coefficients 
$C_1$ and $C_2$. In the leading logarithmic approximation   
we have:
\begin{equation}
C_\pm = C_2 \pm C_1, \quad  C_\pm(\mu)=\left(\frac{\alpha (M_W)}{\alpha (\mu)}\right)^{\frac{\gamma^{(0)}_\pm}{2\beta_0}} C_\pm(M_W), \quad  
C_\pm(M_W) = 1~~,  
\end{equation}
where $\beta_0 = (11 N - 2 f)/{3}  = 23/3$ and 
$\gamma^{(0)}_\pm = \pm  6 (1 \pm N)/ N $. The coefficients to
NLO precision can be found in \cite{WilsonNLL}. 

Because there is another short-distance scale, the bottom quark mass,
the operator product of two $\Delta B = 1$  operators 
can be expanded in inverse powers of the bottom quark mass scale in 
terms of local $\Delta B = 2$ operators: 
\bea
\Gamma_{21} &=& 
\frac{1}{2 M_{B_d}} 
\langle \bar{B}_d | {\rm Im} \, i \int d^4x T{\cal H}^{\Delta B = 1}_{eff}(x) 
{\cal H}^{\Delta B = 1}_{eff}(0) | B_d \rangle \\ 
&=& 
\sum_n\frac{E_n}{m_b^n}\,   \langle \bar{B}_d | {\cal O}_n^{\Delta B=2}(0) | B_d \rangle ~~.
\eea 
These  matching equations fix the values of the $\Delta B = 2$ 
Wilson coefficients
$E_n$. The corresponding four quark operators ${\cal O}_n$ are 
the following: The operators $Q$ and $Q_S$, 
\bea
Q &=& (\overline{b}_id_i)_{V-A} (\overline{b}_jd_j)_{V-A} ,\\ 
Q_S &=& (\overline{b}_id_i)_{S-P} (\overline{b}_jd_j)_{S-P} , 
\eea
represent  the leading order contributions. 
Their matrix elements are given in terms of the bag parameters, $B$ and $B_S$,
the mass of the $B_d$ meson $M_{B_d}$, and its decay constant $f_{B_d}$: 
\bea
\langle  \bar  B_d | Q | B_d 
\rangle  &=& f_{B_d}^2 M_{B_d}^2 2 \frac{N+1}{N} B , 
\\
\langle  \bar  B_d | Q_S | B_d \rangle             &=& - f_{B_d}^2 M_{B_d}^2 
     \frac{M_{B_d}^2}{(\bar{m}_b +\bar{m_d})^2} \frac{2N-1}{N} B_S~~.
\eea
In the naive factorization approximation, $B$ and $B_S$ are fixed by 
$B=B_S=1$. Reliable lattice calculations for
$B$ and $B_S$ are already available \cite{jlqcd}.
We note that to NLO precision one has to distinguish 
between the pole mass $m_b$ and the running quantity 
\beq
\bar m_b (\mu) = m_b \left[ 1 - \frac{\alpha_s}{\pi} \,\,  
\left( ln \frac{\mu^2}{m_b^2} + \frac{4}{3} \right) \right]
\eeq
using the $\overline{MS}$ scheme.
 
The $1/m_b$ corrections are given by the operators
\bea
R_1 &=&\frac{m_d}{m_b}(\bar b_id_i)_{S-P}(\bar b_jd_j)_{S+P} , \\
R_2 &=&\frac{1}{m^2_b}(\bar b_i {\overleftarrow D}_{\!\rho}
\gamma^\mu(1-\gamma_5)
D^\rho d_i)( \bar b_j\gamma_\mu(1-\gamma_5)d_j) , \\
R_3 &=&\frac{1}{m^2_b}(\bar b_i{\overleftarrow D}_{\!\rho}
(1-\gamma_5)D^\rho d_i) (\bar b_j(1-\gamma_5)d_j) , \\
R_4 &=&\frac{1}{m_b}(\bar b_i(1-\gamma_5)iD_\mu d_i)
(\bar b_j\gamma^\mu(1-\gamma_5)d_j),\\
R_0 &=& Q_S + \frac{1}{2} \, Q + \tilde Q_S, 
\label{rrt1}
\eea 
where $\tilde Q_S$ has the ``interchanged'' color structure 
as compared to $Q_S$.
There are also ``color-interchanged'' operators 
$\tilde{R}_i$ and $\tilde{Q}$ corresponding to $R_i$ and $Q$.
We note that these $1/m_b$ operators are not independent, the
relations between them are in fact the equations of motion.

The matrix elements of these operators
within the $B_s$--$\bar B_s$ system were estimated
in \cite{BBD} using naive factorization, which means that 
all the corresponding bag factors were set to $1$.  
For the $B_d$--$\bar B_d$ system the analogous
results are:  
\bea
\langle  \bar B_d | R_0 | B_d \rangle  &=& 
  f^2_{B_d}M^2_{B_d} \left( \frac{N + 1}{N} \right)
  \left( 1-\frac{M^2_{B_d}}{m_b^2} \right) ~, \label{r0} \\
  \langle  \bar  B_d | R_1 | B_d \rangle  &=& 
  f_{B_d}^2 M_{B_d}^2 \frac{m_d}{m_b} \frac{2 N +1}{N} = 0 ~, \\
  \langle  \bar  B_d | \tilde{R}_1 | B_d \rangle  &=& 
  f_{B_d}^2 M_{B_d}^2 \frac{m_d}{m_b} \frac{N + 2}{N} = 0 ~, \\
\langle  \bar  B_d | R_2 | B_d \rangle   &=& 
  f_{B_d}^2 M_{B_d}^2  
  \left( \frac{M_{B_d}^2}{m_b^2} - 1 \right)\frac{1 - N}{N}
\, = \, - \langle  \bar  B_d | \tilde{R}_2 | B_d \rangle  ~,\\
\langle  \bar  B_d | R_3 | B_d \rangle   &=& 
  f_{B_d}^2 M_{B_d}^2 
  \left( \frac{M_{B_d}^2}{m_b^2} - 1 \right)\frac{2 N + 1}{2 N}~,\\
\langle  \bar  B_d | \tilde{R}_3 | B_d \rangle   &=& 
  f_{B_d}^2 M_{B_d}^2 
  \left( \frac{M_{B_d}^2}{m_b^2} - 1 \right)\frac{N + 2}{2 N}~,\\
\langle  \bar  B_d | R_4 | B_d \rangle   &=& 
   - f_{B_d}^2 M_{B_d}^2  
  \left( \frac{M_{B_d}^2}{m_b^2} - 1 \right)~, \\
\langle  \bar  B_d | \tilde{R}_4 | B_d \rangle   &=& 
   - f_{B_d}^2 M_{B_d}^2  
  \left( \frac{M_{B_d}^2}{m_b^2} - 1 \right)\frac{1}{N}~~.
\label{r4bar}
\eea 
We neglect terms proportional to $m_d/m_b$; the other terms 
proportional to   $ (M_{B_d}^2 / m_b^2)  - 1$  are of order
$\Lambda_{QCD}/m_b$.

In the matrix elements 
$\langle  R_i\rangle $ (eqs. (\ref{r0})--(\ref{r4bar})), 
we use the pole mass $m_b$. 
There is a subtlety involved here:
as discussed in \cite{BBGLN}, there are terms of order
$\alpha_s$ and of leading power in $m_b$ in the matrix element 
of $R_0$ to NLO precision. 
In view of the relation (\ref{rrt1}), it is not surprising  that there
are such terms. In the scheme -- which was used in \cite{BBGLN} and which
is  also used here -- these terms are subtracted in the matrix element 
$\langle  R_0\rangle $ while taking into account  the leading NLO 
contribution. 
Then the $\langle  R_0\rangle $ matrix element is still 
of a subleading nature. 
The specific subtraction scheme for the factorized
matrix elements $\langle  R_i\rangle $  corresponds to the use
of the pole mass in eqs. (\ref{r0})--(\ref{r4bar}).
Of course this specific choice  for the matrix elements
has to be taken into account if the NLO results are combined
with a lattice calculation of the $\langle  R_i\rangle $.

There is an additional remark in order.
We estimate $\Gamma_{21}$ by the cut of the partonic diagrams.
The underlying assumption of local quark-hadron duality
can be verified in the $B_s$--$\bar B_s$ system, in the 
simultaneous limit of large $N$  and of small velocity 
\cite{aleksan}, 
therefore one expects no large duality violations.
In the $B_d$--$\bar{B}_d$ system the small velocity 
argument fails since the $u\bar u$, $u\bar c$ and $c\bar u$ 
intermediate states contribute significantly, and
the larger number of light
intermediate states leads to a larger energy release.
We follow ref. \cite{BBD} and make the assumption that the duality
violations in the $B_d-\bar B_d$ system are also
not larger than $10\%$.
In order to test this assumption one should include all corrections up to 
that accuracy.

\subsection{Analytical results} 
\label{corr}

In this section, we present an analytic expression 
for $\Gamma_{21}$ including $1/m_b$, penguin and part of the NLO
corrections.
If one takes into account the error inherent in the naive
factorization approach to the matrix elements of the subleading
operators $R$, it seems to be a reasonable approximation to keep
at least all terms up to an accuracy of $10^{-2}~ 
\Gamma_{21}^{leading}$.
We keep also  higher order terms in order to check the accuracy of our 
approximation. 

In the effective theory of the $\Delta B =2$ transitions
the matrix elements of the 
$1/m_b$ operators ($R$) are formally suppressed by a factor 
of the order of $0.1$ with respect to
those of the leading operators $Q$ and $Q_S$.
The natural variable $z=m_c^2/m_b^2$ also formally introduces a 
suppression factor of approximately $0.1$. 
The NLO contribution has formally an extra suppression
factor $(\alpha_s/4\pi)$ of order $0.01$. 
Within the effective
theory of the $\Delta B = 1$ Hamiltonian, 
the combination $K' = C^{peng} C^{dom}$ and $K'' = C^{peng} C^{peng}$
are suppressed by almost a  factor 0.01 and $10^{-4}$ respectively, 
with respect to the combination $K = C^{dom} C^{dom}$, where $C^{peng}$
denotes the Wilson coefficients of the penguin
operators $Q_3...Q_6$ and $C^{dom}$ that of the dominating operators 
$Q^{qq'}_{1 \, (2)}$. The contribution due to $K''$ 
therefore can be safely neglected.
Schematically our analytical result for $\Gamma_{21}$ has  
the following form:
\bea
\Gamma_{21} &=& \, K\,  \langle {\cal Q}\rangle  
\label{leading}\\
  &+& \, K \, \langle  R\rangle  \, (O(1)+O(z)+O(z^2)+\{O(z^3)\})
\label{1/mb}\\
  &+& \, K' \, \langle {\cal Q}\rangle  \, (O(1)+O(z)+\{O(z^2)\})
\label{peng1}\\
  &+& \, K' \,  \langle  R\rangle  \,(O(1)+\{O(z^2)\}
\label{peng2}\\
  &+& \, \alpha_s/(4\pi) \,\,  K \, \langle {\cal Q}\rangle  \, O(1)~~,
\label{alpha2}
\eea
where ${\cal Q}$ represents the leading order operators
$Q$ and $Q_S$.
The terms inside the curly brackets are the ones
that we calculate only to estimate the errors.
In the presentation of the results the following 
combinations of the Wilson coefficients are used: 
\bea
K_1 &=& 3 C_1^2 + 2 C_1 C_2,
\quad K_2 = C_2^2 , \quad K_3 = C_1^2, \quad K_4 = C_1 C_2\\
K_1^\prime &=& 2 (3 C_1 C_3 + C_1 C_4 + C_2 C_3 ), ~~
K_2^\prime = 2 C_2 C_4, \\
K_3^\prime &=& 2 ( 3 C_1 C_5 + C_1 C_6 + C_2 C_5 + C_2 C_6 ), 
\eea  
and the common factor of $[-G_F^2 m_b^2/ (24 \pi  M_{B_d})]$ is
implicit in the following equations (\ref{g-leading}), (\ref{g-1/mb}),
(\ref{g-peng}), (\ref{g-alpha}).

In the leading log approximation we calculate the $Q^{qq'}_1$ and the 
$Q^{qq'}_2$  contributions to $\Gamma_{21}$.
By extracting the absorptive parts of  
the $c\bar{c},u\bar{c},c\bar{u}$ and $u\bar{u}$ 
intermediate states, we can find the off-diagonal element.
For this leading  contribution (\ref{leading}),
after replacing $V_{ub}^* V_{ud} $ by the unitarity relation,
we get to all orders in $z$: 
\bea
\Gamma^{leading}_{21} & = &
 (V_{tb}^* V_{td} )^2 
          \left[  \left( K_1 + \frac{1}{2}K_2 \right)\langle  Q\rangle  
           + ( K_1 - K_2 ) \langle  Q_S\rangle  \right] \nn \\
 &+& ( V_{cb}^*V_{cd})(V_{tb}^*V_{td} ) \Bigl[
           ( 3 z ( K_1 + K_2 ) - 3 z^2 K_2 - z^3 (K_1-K_2))\langle  Q\rangle   \nn \\
 & & \quad \quad \quad \quad
           + (6 z^2 ( K_1 - K_2 ) - 4 z^3 ( K_1 - K_2 ) ) \langle  Q_S\rangle \Bigr]  \nn \\
 &+& (V_{cb}^* V_{cd} )^2 \Bigl\{ \sqrt{1-4 z} \left[ \left( 
        K_1 + \frac{1}{2} K_2 \right) -  
      z (K_1 +  2 K_2) \right] \langle  Q\rangle  \nn \\
 & & \quad \quad \quad \quad + \sqrt{1-4 z} (1+2z)(K_1-K_2) 
                \langle  Q_s\rangle  \nn \\
 & & \quad \quad \quad \quad + 
\left( (1-z)^2 \left[ - (2 K_1 + K_2) + z (K_2 - K_1) \right] + 
 \left[ K_1 + \frac{1}{2} K_2 \right] \right) \langle  Q\rangle  \nn \\
 & & \quad \quad \quad \quad +
\left[ (1-z)^2 (2+4 z)(K_2 - K_1) - (K_2-K_1) \right] \langle  
Q_S\rangle \Bigr\} ~~.  
\label{g-leading}
\eea
 This result, (\ref{g-leading}), 
confirms the findings of \cite{SM2}.

In the equations  
(\ref{g-1/mb}),(\ref{g-1/mb2}), (\ref{g-peng}), and (\ref{g-peng2})
we give  the penguin and  $1/m_b$ contributions. 
The coefficients of $(V_{tb}^* V_{td})^2$ in these results can be
checked with the $z \to 0$ limit of the results in the literature
for the $B_s$ syatem.
In this sense our results are consistent with the findings  
in the $B_s$ system given in \cite{BBD}.

The $1/m_b$ corrections to the operators $Q^{qq'}_1$ and $Q^{qq'}_2$
give [see the term (\ref{1/mb})]
\bea
\Gamma^{1/m_b}_{21} &=& (V_{tb}^* V_{td})^2 
\left[ - 2 \left( K_1 - \frac{1}{2} K_2 \right) \langle  R_2\rangle
- 2 K_1 \langle  R_1\rangle  + 2 K_2 \langle  R_4\rangle  \right]  \nn \\
&+& ( V_{cb}^*V_{cd})(V_{tb}^*V_{td} ) 
    \Bigl[ - 12 z^2 K_1 ( \langle  R_1\rangle  - 2 \langle  R_3\rangle  ) \nn \\
& & ~~~~~~~~~~~~~~~~~~~~
    + 6 z^2 K_2 ( \langle  R_2\rangle  + 4 \langle  R_3\rangle  
                                   + 2 \langle  R_4\rangle ) \Bigr] \nn \\
&+& \Bigl\{ K \langle  R \rangle    O(z^3) \Bigr\}.
\label{g-1/mb}
\eea
The term in curly brackets in (\ref{g-1/mb}) can be written as
\bea 
\{ ... \} & = & ( V_{cb}^*V_{cd})(V_{tb}^*V_{td} ) \times \nn \\
& &  [ 4 z^3 K_1 ( 2 \langle  R_1\rangle - \langle  R_2\rangle 
                  -6 \langle  R_3\rangle ) 
    - 4 z^3 K_2 (  \langle  R_2\rangle + 6 \langle  R_3\rangle 
                  +2 \langle  R_4 \rangle ) ]\nn \\
&+& (V_{cb}^* V_{cd})^2 \times \nn \\
& & [ 12 z^3 K_1 ( 2 \langle  R_1\rangle - \langle  R_2\rangle 
                  -6 \langle  R_3\rangle ) 
    -12 z^3 K_2 (  \langle  R_2\rangle + 6 \langle  R_3\rangle 
                  +2 \langle  R_4 \rangle ) ]~~. 
\label{g-1/mb2}
\eea 

The penguin contributions [terms (\ref{peng1}), (\ref{peng2})] are
\bea
\Gamma^{peng}_{21} &=& (V_{tb}^* V_{td})^2 
 \Bigl[ \left( K_1^\prime + \frac{1}{2} K_2^\prime \right) \langle  Q\rangle  
  + ( K_1^\prime - K_2^\prime )\langle  Q_S\rangle  \nn \\
& & ~~~~~~~~~ +   
    ( - 2 \langle  R_2\rangle  - 2 \langle  R_1\rangle  ) K_1^\prime 
  + ( \langle  R_2\rangle  + 2 \langle  R_4\rangle  ) K_2^\prime \Bigr] \nn \\
&+& ( V_{cb}^*V_{cd})(V_{tb}^*V_{td} ) 
    ( 3z K_1^\prime + 3z K_2^\prime 
              - 3z K_3^\prime )\langle  Q\rangle   
              \nn \\
   & &+ \bigl\{ K' \, \langle {\cal Q}\rangle  \, O(z^2)
  + \, K' \, \langle  R\rangle  \, O(z) \bigr\} ~~,
\label{g-peng}
\eea
where the terms in curly brackets (and the lower order ones) 
may be written as
\bea
\{ ... \} & = & 
 ( V_{cb}^*V_{cd})(V_{tb}^*V_{td} ) \times   \nn \\ 
  &  & \Bigl[~~  
        ( - 3z^2  K_2^\prime +6z^2 K_3^\prime )\langle  Q\rangle   
     +( 6 z^2 K_1^\prime -6z^2 K_2^\prime )\langle  Q_S\rangle  \nn \\
 & &  -(12z^2 \langle R_1 \rangle -24z^2 \langle R_3 \rangle) 
        K_1^\prime  \nn \\
 & & + (6z^2 \langle R_2 \rangle + 24z^2 \langle R_3 \rangle 
        + 12z^2 \langle R_4 \rangle) K_2^\prime
        + 4 z^2 \langle R_2 \rangle K_3^\prime ~~\Bigr]~.
\label{g-peng2}
\eea

The NLO QCD correction 
$\Gamma^{NLO}_{21} = \alpha_s/(4\pi) \,\,  K \, \langle {\cal Q}\rangle $ 
[term (\ref{alpha2})] is found from \cite{BBGLN}
by taking the limit $z\rightarrow 0$ of their results\footnote{
We add only the leading contribution of the NLO QCD corrections 
for the term 
$(V_{tb}^* V_{td} )^2$. The leading terms of the contributions 
for the terms  
$( V_{cb}^*V_{cd})(V_{td}^*V_{td} )$ and $( V_{cb}^*V_{cd})^2 $ cancel out 
through the GIM mechanism. }:
\bea
\Gamma^{NLO}_{21} &=& \frac{\alpha_s(m_b)}{4 \pi }
     (V_{tb}^* V_{td} )^2 
        \Bigl\{  \left[  \frac{109}{6} K_3 
      - \frac{248}{9} K_4 
      - \left( \frac{\pi^2}{3} + \frac{157}{18} \right) K_2 \right] 
        \langle  Q\rangle  \nn \\
      & &\quad \quad  + \left[ \left( 10 K_3 + \frac{20}{3} K_4 + 
        \frac{8}{3} K_2 \right)  
        {\rm ln} \left(\frac{\mu_2}{m_b} \right)  - 
        (34 K_4 + 10 K_2) {\rm ln} \left(\frac{\mu_1}{m_b} \right) \right] 
        \langle  Q\rangle  \nn \\
   & & \quad \quad - \left[ \frac{40}{3} K_3 + \frac{248}{9} K_4 
        - \left( \frac{8 \pi^2 }{3 } - \frac{128}{9} \right) K_2 \right] 
        \langle  Q_S\rangle  \nn \\
   & & \quad \quad + \left[ \left(32 K_3 - \frac{64}{3} K_4 + 
                \frac{32}{3} K_2 \right) 
        {\rm ln} \left(\frac{\mu_2}{m_b} \right) - 
        (16 K_4 + 16 K_2) {\rm ln} \left(\frac{\mu_1}{m_b} \right) 
                \right] \langle  Q_S\rangle  \nn \\
& & \quad \quad - \left[ \frac{2}{27} + 
        \frac{2}{9} {\rm ln} \left(\frac{\mu_1}{m_b} \right) + 
        \frac{1}{3} \frac{C_8}{C_2}\right] K_2 
        (\langle  Q\rangle  - 8\langle  Q_S\rangle )\Bigr\} ~.
\label{g-alpha}
\eea
The explicit $\mu_1$ and $\mu_2$ dependence 
in (\ref{g-alpha}) cancels against the $\mu$ dependence of the Wilson 
coefficients of the hamiltonian ${\cal H}_{eff}^{\Delta B =1}$ (\ref{HHH}) and 
the $\mu$ dependence of the matrix elements of the $\Delta B =2$ operators 
at the order
 in $\alpha_s$ we take into account. For a 
proper matching with lattice 
evaluations of these matrix elements it is important to note that
the results in (\ref{g-alpha}) are based on the NDR scheme, with
the choice of $\gamma_5$ and the evanescent operators as 
given in eqs. (13)--(15) of \cite{BBGLN}.

The net $\Gamma_{21}$ is   
\beq
\Gamma_{21} = \Gamma_{21}^{leading} + \Gamma_{21}^{1/m_b} +
\Gamma_{21}^{peng} + \Gamma_{21}^{NLO}~~,
\label{netgamma}
\eeq
with the implicit multiplicative factor of  
$[-G_F^2 m_b^2/ (24 \pi  M_{B_d})]$.

\subsection{Numerical results}
\label{numerical}

Let us now calculate the numerical value of $\dg$. From 
eq. (\ref{Dapprox}), $\Delta \Gamma_d $ can be approximately 
written as 
\bea
\Delta \Gamma_d \approx - 2 |M_{21}| {\rm Re} \frac{\Gamma_{21}}{M_{21}}
       = -\Delta m~  {\rm Re}\frac{\Gamma_{21}}{M_{21}}~.
\eea 
where $M_{21}$ [see eq. (\ref{m12})] is given by
\bea
M_{21} = \frac{G_F^2 M_W^2 \eta_B }{(4 \pi)^2 (2 M_{B_d} )
      } (V_{tb}^* V_{td})^2 
       S_0(x_t) \langle  Q\rangle   .
\eea
Here $x_t = \bar{m}_t^2/M_W^2 $, $\eta_B $ is the QCD correction 
factor and $S_0$ is the Inami--Lim function:
\bea
S_0(x) = x \left( \frac{1}{4} + \frac{9}{4(1-x)}-\frac{3}{2(1-x)^2} \right)
      - \frac{3}{2}\left( \frac{x}{1-x} \right)^3 \log x ~~.
\eea
Using the results obtained in the previous section, we can
write down the width difference (normalized to the average width)
in the form
\bea
\left(\frac{\Delta \Gamma}{\Gamma }\right)_{B_d} &=& 
\left(\frac{\Delta m}{\Gamma }\right)_d  {\cal K} \times \nn \\
 & & 
\left[  G^{tt} + \frac{5}{8}\frac{B_S~M_{B_d}^2}{B~\bar{m}_b^2} G_S^{tt} 
               + \frac{3}{8}\frac{1}{B} G_{1/m}^{tt} \right. \nn \\
 & &  + {\rm Re}\left(\frac{V_{cb}^* V_{cd}}{V_{tb}^* V_{td}}\right)
        \cdot \left(
        G^{ct} + \frac{5}{8}\frac{B_S~M_{B_d}^2}{B~\bar{m}_b^2} G_S^{ct} 
              + \frac{3}{8}\frac{1}{B} \{G_{1/m}^{ct}\} \right) \nn \\
& & + \left. 
        {\rm Re}\left(\frac{V_{cb}^* V_{cd}}{V_{tb}^* V_{td}}\right)^2  \cdot
        \left( G^{cc} +
        \frac{5}{8}\frac{B_S~M_{B_d}^2}{B~\bar{m}_b^2} G_S^{cc} \right)
        \right] ~~.
\label{g-form}
\eea
The superscripts $\{ tt, ct, cc \}$ correspond to the terms
in the expression for $\dg$ (\ref{netgamma}) that involve the CKM
factors $\{ (V_{tb}^* V_{td})^2, (V_{cb}^* V_{cd} )(V_{tb}^*V_{td}),
(V_{cb}^*V_{cd})^2 \}$ respectively. 
The subscript $S$ denotes the contribution
from the  operator $Q_S$, and the subscript $1/m$ denotes the terms
that give the $1/m_b$ corrections. 
The normalizing factor 
${\cal K} \equiv (4\pi m_b^2)/(3 M_W^2 \eta_{B} S_0(x_t))$
and the value of $(\Delta m / \Gamma)_d$ may be taken from experiments:
$x_d \equiv (\Delta m / \Gamma)_d = 0.73 \pm 0.03$
\cite{pdg}. 
The form of eq.
(\ref{g-form}) can bring out important features of the dependence
of $\dg$ on various parameters, as we shall see below. 
This representation also has the advantage that
within the leading term the CKM dependence cancels out and 
the value of $x_d$ is available from experiments.

A remark about the penguin contributions is in order. 
We only include the  interference of the penguin operators
$C_3...C_6$ with the leading operators $C_1$ and $C_2$. 
At the NLO, this
approximation can be made consistent (in the sense of scheme
independence) by counting the Wilson coefficients
$C_3...C_6$ as of order $\alpha_s$. These Wilson coefficients
are modified at NLO through the mixing of $C_1$ and $C_2$ into 
$C_3...C_6$. 
For $C_1$ and $C_2$ we use the complete NLO values. 
Since the contribution due to $C_8$ starts only at the NLO level, 
we only have to use the LO value for that Wilson coefficient.
We stress that if one uses the consistent NLO
approximation just described, the corresponding  LO approximation 
includes no penguin contributions and uses 
the LO values for $C_1$ and $C_2$.

The choice of the $b$-quark mass at LO is ambiguous 
(it may be taken to be the pole mass or the running mass
at one or two loop level); we take it to be the running mass in
the MS scheme to leading order in $m_b$.

We use the following values of parameters to estimate $\dg$:
\bea
M_{B_d} = 5.28 ~{\rm GeV} ~,~
m_b = 4.8 ~{\rm GeV} ~,~
m_c = 1.4 ~{\rm GeV}~,~ \nn \\ 
\bar{m}_b(m_b) = 4.4  ~{\rm GeV} ~,~ 
\bar{m}_t(m_b) = 167  ~{\rm GeV} ~. 
\eea
To the NLO precision [we use here
the NDR scheme to get $\eta_B(m_b) = 0.846$ and include the
NLO Wilson coefficients \cite{bbl} 
and the corrections computed in eqs.~(\ref{g-1/mb}),(\ref{g-peng})], 
we get (in units of $10^{-3}$)
\bea
\left(\frac{\Delta \Gamma}{\Gamma }\right)_{B_d} & = & 
  0.73 + 4.93 \frac{B_S}{B }  -1.38 \frac{1}{B} \nn \\  
 & &  -\frac{\cos\beta }{R_t } \left(1.07 + 0.29 \frac{B_S}{B} + 
      0.02 \frac{1}{B}  
           + \left\{-0.005 - 0.021 \frac{B_S}{B} 
                    - 0.003 \frac{1}{B} \right\} \right) \nn \\
 & &  + \frac{\cos2\beta }{R_t^2}  \left(0.02 - 0.06\frac{B_S}{B } 
        + \left\{- 0.01 \frac{1}{B} \right\}              \right) ~~.
\label{dg-nlo}
\eea

Let us perform a conservative estimate of the error on the
value of $\dg/\Gamma_d$ that we obtain here. The errors 
arise from the uncertainties in the values of the CKM parameters, 
the bag parameters and the mass of the $b$ quark.
There are also errors from the scale dependence, the breaking of
the naive factorization approximation, and the neglected
higher order terms in the $z$ expansion.

In the SM, we have
\beq
\cos\beta/R_t = 1.03 \pm 0.08  \quad , \quad 
\cos 2\beta/R_t^2 = 0.87 \pm 0.15  ~~,
\label{km-error}
\eeq
where we have taken the values of the CKM parameters from the
global fit \cite{ckm-fit}.
The leading term on the first line in (\ref{dg-nlo}) is independent
of the CKM elements.
The quantity $\cos\beta/R_t$ is known to an accuracy of about 
10\% and appears in (\ref{dg-nlo}) with a coefficient $\sim 0.3$
relative to the leading term. The quantity $\cos 2\beta/R_t^2$,
although known to only about 20\%, appears with a very small 
coefficient ($\sim 10^{-2}$) as compared to the leading term in 
(\ref{dg-nlo}). 
The net error due to the uncertainty in the CKM elements is thus 
approximately only 3\%, i.e. about $\pm 0.1 \times 10^{-3}$.

\begin{table}[tbh]
\begin{center}
\begin{tabular}{|c|c|c|c|c|}
\hline
        & LO     & A     & B     & C \\ \hline
 $B$    & 1.0   & 0.90   & 0.83   & 1.0 \\ 
 $B_S$  & 1.0   & 0.75  & 0.84   & 1.0 \\ \hline 
$\Delta \Gamma_d/\Gamma_d $ & $ 6.3 \times 10^{-3} $ & 
     $1.9 \times 10^{-3} $ & 
     $ 2.6 \times 10^{-3} $ & $ 2.8 \times 10^{-3} $ \\ \hline 
\end{tabular}
\caption{The numerical value of $\dg/\Gamma$ for different values
of the bag parameters. The column LO (C) shows the leading 
(next-to-leading) order result with factorization, i.e.
$B = B_S = 1$. The values of the bag factors in column A are 
taken from \cite{BBGLN} and the ones in column B  
from the (preliminary) results in an unquenched ($N_f=2$)
lattice calculation by the JLQCD collaboration \cite{jlqcd}.}
\label{diff-bag}
\end{center}
\end{table}

We estimate the effect of the uncertainties in the bag factors by
computing (\ref{dg-nlo}) with three sets of values of the bag 
parameters. The numerical results are as shown in 
Table~\ref{diff-bag}. From the table, 
and using the uncertainties on the values of
the bag parameters as given in \cite{lenz}, we conservatively 
estimate the corresponding uncertainty in the value of 
$\dg/\Gamma_d$ due to bag 
factors to be approximately $\pm 0.5 \times 10^{-3}$.
The uncertainty in the value of $\bar{m_b} = 4.4 \pm 0.2$ also 
leads to an error of $\pm 0.5 \times 10^{-3}$. 
The uncertainty due to the scale $\mu_1$ dependence is estimated to be
$^{+0.5}_{-1.2} \times 10^{-3}$ (where $\mu_1$ is varied between
$2 m_b$ and $m_b/2$ following the common convention). 
The error due to the input value of $x_d$ is $0.1 \times 10^{-3}$.

The errors due to the breaking of the naive factorization assumption
(which was made in the calculation of the matrix elements of the
$1/m_b$ operators) are hard to quantify. Assuming an error of 
30\% in the $R$ matrix elements (as in \cite{lenz}), we estimate 
the error due to this source to be $\pm 0.3 \times 10^{-3}$.

Table~\ref{diff-bag} also gives the LO value of $\dg/\Gamma_d$ in the
factorization approximation.
We observe that the NLO corrections significantly decrease
the value of $\dg/\Gamma_d$ as computed at LO, and that there effectively 
is no real $(\alpha_s/4\pi)$ suppression of the NLO contribution, 
as one naively expects. Therefore higher-order terms in the $z$ 
expansion become important. 
While we estimate the error due the $z$ expansion 
in the $1/m_b$ and the penguin contributions
from the terms in curly  brackets in (\ref{dg-nlo}) to be  
less than $\pm 0.05 \times 10^{-3}$,
the issue of higher order terms in the NLO contribution (\ref{g-alpha})
is more subtle. 
We can write $\Gamma_{21}$ in the form (see Sec.~\ref{upper-bd}
for details)
\bea
\Gamma_{21}(B_d) & = &  -{\cal N}  
\Bigl[ (V_{cb}^* V_{cd})^2 [f(z,z)-2f(z,0)+f(0,0)]  \nn \\ 
 & &  + 2 (V_{cb}^* V_{cd})(V_{tb}^* V_{td}) [ f(0,0) -f(z,0)] 
                        + (V_{tb}^* V_{td})^2 f(0,0) \Bigr]~~.
\label{bd-g12-extra}
\eea
Here in (\ref{g-alpha}) we have included 
the complete NLO coefficient of $(V_{tb}^* V_{td})^2$, 
which includes all the terms
of the order $z^0$ in the hadronic matrix elements $f$.
However, in order to calculate the corrections due to higher order
terms in $z$, a complete NLO calculation is necessary. The 
contribution of these terms to $[(f(0,0)-f(z,z))/f(0,0)]$ can
be computed to NLO precision using \cite{BBGLN} to be 
\beq
[(f(0,0)-f(z,z))/f(0,0)]_{NLO} \approx 0.2~~.
\label{assumea}
\eeq
If we estimate the contribution to $[(f(0,0)-f(z,0))/f(0,0)]_{NLO}$
to be also of the same order,
 this results in the 
estimation of the net error in $\dg/\Gamma_d$ due to
these terms to be $\pm 0.8 \times 10^{-3}$.

Our net estimation for the width difference is 
\beq
\left( \frac{\dg}{\Gamma_d} \right)_{B_d} = ( 2.6 ^{+1.2}_{-1.6})
\times 10^{-3}~~.
\label{value}
\eeq
We have taken the central value to be the one obtained from the latest
preliminary (unquenched) results from lattice calculations \cite{jlqcd}.
The dominating theoretical errors are the scale dependence and the terms
in $\Gamma_{21}^{NLO}$ that correspond to the nontrivial $z$ dependence 
of the function $f(z,0)$ in 
(\ref{bd-g12-extra}).
To take care of the latter, a complete NLO calculation is definitely
desirable.

In the above calculations, we have used the expansion of $\Gamma_{21}$
in the form
\barr
\Gamma_{21}(B_d) & = & - {\cal N} (V_{tb}^*V_{td})^2 f(0,0) 
\left[ 1 + {\cal F}_{ct} 
\left(\frac{V_{cb}^* V_{cd}}{V_{tb}^*V_{td}}\right) + {\cal F}_{cc}
\left(\frac{V_{cb}^*V_{cd} }{V_{tb}^*V_{td}}\right)^2 \right] \nonumber \\
 & \equiv & - {\cal N} (V_{tb}^*V_{td})^2 f(0,0) (1 + \delta f_c) ~~,
\label{ct-form}
\earr
where ${\cal F}_{ct} \equiv 2 [f(0,0)-f(z,0)]/f(0,0)$ and
${\cal F}_{cc} \equiv  [f(0,0)- 2 f(z,0) + f(z,z)]/f(0,0)$.

Following the suggestion in \cite{breport}\footnote{We thank Uli Nierste
 for bringing this to our attention.}, we have also performed the
expansion (and the error analysis) in the form
\barr
\Gamma_{21}(B_d)&  = &  - {\cal N} (V_{tb}^*V_{td})^2 f(z,z) 
\left[ 1 + {\cal F}_{ut} 
\left(\frac{V_{ub}^*V_{ud}}{V_{tb}^*V_{td}}\right) + {\cal F}_{uu}
\left(\frac{V_{ub}^*V_{ud}}{V_{tb}^*V_{td}}\right)^2 \right] \nonumber \\
 & \equiv & - {\cal N} (V_{tb}^*V_{td})^2 f(z,z) (1 + \delta f_u) ~~,
\label{ut-form}
\earr
where ${\cal F}_{ut} \equiv 2 [f(z,z)-f(z,0)]/f(z,z)$ and
${\cal F}_{uu} \equiv  [f(0,0)- 2 f(z,0) + f(z,z)]/f(z,z)$.
In this expansion the unknown NLO terms are suppressed by 
small CKM factors.  
This gives the width difference as
\beq
\left( \frac{\dg}{\Gamma_d} \right)_{B_d} = ( 3.0 ^{+0.9}_{-1.4})
\times 10^{-3}~~,
\label{value1}
\eeq
where, as before, we use the latest 
preliminary (unquenched) results from lattice calculations \cite{jlqcd}
for the bag parameters.
The results of both (\ref{value}) and (\ref{value1}) are consistent.
The errors in (\ref{value1}) are smaller, but it should be noted that
in both calculations 
the errors due to NLO terms are based on the assumption on the function
$f(z,0)$ which we stated above.

\section{Measurement of $\dg/\Gamma_d$}
\label{phenom}

It is not possible to find a final state to which the decay of 
$B_d$ involves only one of the decay widths $\gl$ and $\gh$.
Indeed, since the $B_d$--$\bar{B}_d$ mixing phase ($2\beta$)
is large, the CP eigenstates are appreciably different
from the lifetime eigenstates. The decay rate to a CP
eigenstate therefore involves both the lifetimes. The semileptonic
decays are flavor-tagging, and hence also involve both the lifetimes
in equal proportion.

We start by concentrating on the untagged measurements, i.e. the
measurements in which the $(\Delta m t)$ oscillations are
cancelled out. When the production asymmetry between $B_d$ and
$\bar{B}_d$ is zero (as is the case at the $B$ factories), this
corresponds to not having to determine whether the decaying meson
was $B_d$ or $\bar{B}_d$. Restricting ourselves to untagged
measurements is a way of getting rid of tagging
inefficiencies and mistagging problems. 
At hadronic machines, a handle on the production asymmetry
between $B_d$ and $\bar{B}_d$ is necessary. 

In this section, we show explicitly that the time measurements of the decay 
of an untagged $B_d$ to a single final state can only be sensitive to
quadratic terms in $\dg/\Gamma_d$. This would imply that, for determining 
$\dg/\Gamma_d$ using only one final state, the accuracy of the
measurement needs to be $(\dg/\Gamma_d)^2 \sim  10^{-5}$.
This indicates the necessity of combining measurements from two
different final states to be sensitive to a quantity
linear in $\dg/\Gamma_d \sim 0.3 \times 10^{-2}$.
We discuss three pairs of candidate channels for achieving
this task. 
Finally, we point out the extent of 
systematic error in the conventional measurement of $\beta$
due to the neglect of the width difference, and show how
the tagged $B_d \to J/\psi K_S$ mode can also measure
$\dg/\Gamma_d$ by itself.

\subsection{Quadratic sensitivity to $\dg/\Gamma_d$ of untagged
measurements}
\label{quad}

It is ``common wisdom'' that the time measurements, in general, are
sensitive only quadratically to $\dg/\Gamma_d$. Specific calculations
(e.g. see \cite{dfn}) also get results that can be clearly seen to
obey this rule. Here, we give an explicit derivation of the 
general statement, pointing out the exact conditions under which 
the above statement is valid. Ways of getting around these conditions
lead us to the decay modes that can provide measurements sensitive
linearly to $\dg/\Gamma_d$.

The non-oscillating part of the proper time distribution of the 
decay of $B_d$ can be written in the most general form as
\beq
f(t) = \frac{1}{2} \left[ (1+b) e^{-\gl t} + (1-b) e^{-\gh t} \right]~~.
\label{ft}
\eeq
The non-oscillating part can also be looked upon as the untagged
measurement.

For an isotropic decay,
the only information available from the experiment is the time $t$.
This information may be {\it completely} encoded in terms of the
(infinitely many) time moments 
\beq
\langle  t^n \rangle  \equiv \frac{\int t^n f(t) dt}{\int f(t) dt}
~~.
\label{tn-def}
\eeq
Expanding in powers of $\dg/\Gamma_d$, we get
\beq
\langle  t^n \rangle  =  \frac{n!}{(\Gamma_d)^n} 
\left[ 1 - \frac{n~b}{2} \frac{\dg}{\Gamma_d} 
+ {\cal O} \left[ (\dg/\Gamma_d)^2 \right] \right] ~.
\label{tn-exp}
\eeq
Defining the {\it effective untagged lifetime} as 
$\tau_b \equiv \frac{1}{\Gamma_d} \left( 1 - \frac{b}{2}
\frac{\dg}{\Gamma_d} \right)$, all the available information 
(\ref{tn-def}) is encoded in
\beq
\langle  t^n \rangle  = n ! (\tau_b)^n \left[ 1 
+ {\cal O} \left[ (\dg/\Gamma_d)^2 \right] \right] ~.
\label{tau-n}
\eeq
Thus, when the accuracy of the lifetime measurement is less than 
$(\dg/\Gamma_d)^2$,
only the combination $\tau_b$ of $\Gamma_d, \dg$ and $b$
may be measured through a single final state.
This measurement is insensitive to $b$ (to this order) and
hence incapable of even discerning the presence of two distinct
lifetimes ($b=0$ and $b=1$ would correspond to the presence of
only a single lifetime involved in the decay.)
In particular, in order to determine $\dg/\Gamma_d$, the
lifetime measurement through the semileptonic decay needs to be
more accurate than $(\dg/\Gamma_d)^2 \sim 10^{-5}$.
This task is beyond the capacity of the currently planned 
experiments.

Combining time measurements from two different final states,
however, can enable us to measure quantities linear in 
$\dg/\Gamma_d$. Indeed, for two final states with different
values $b$ (say $b_1$ and $b_2$), we can measure
\beq
\frac{\tau_{b_1}}{\tau_{b_2}} = 
1 + \frac{b_2 - b_1}{2} \frac{\dg}{\Gamma_d} +
{\cal O} \left[ (\dg/\Gamma_d)^2 \right] ~~.
\label{t1-t2}
\eeq
In the next subsections, we discuss pairs of decay channels that
can measure this quantity (\ref{t1-t2}) that is linear in
$\dg/\Gamma_d$.

\subsection{Decay widths in semileptonic and CP-specific channels}
\label{semilep}

Let us first develop the formalism that will be applicable
for all the decays that we shall consider below.
When the width difference is taken into account, the decay rate
of an initial $B_d$ to a final state $f$ is given as follows.
Let $A_f \equiv \langle  f | B_d \rangle ,~\bar{A}_f \equiv 
\langle  f | \bar{B}_d \rangle  $, and 
\beq
\lambda_f \equiv \frac{q}{p} \frac{\bar{A}_f}{A_f}~,
\label{lambda-def}
\eeq
where $p$ and $q$ are as defined in (\ref{p-q}).
Using the CP-violating parameter $\delta^d$ as defined in
(\ref{delta-def}), we get 
\beq
\left|\frac{q}{p} \right| = \sqrt{\frac{1-\delta^d}{1+\delta^d}}
\approx 1 - \delta^d~~.
\label{q/p}
\eeq
The approximation here is valid since we have
$|\delta^d | \sim  |\dg / \Delta m_d| 
\lsim 10^{-2}$. Henceforth, we shall only consider terms linear in
$\delta^d$.

The decay rate of an initial tagged $B_d$ or $\bar{B}_d$
to a final state $f$ is given by
\cite{dfn}:
\barr
\!\!\! \Gamma(B_d(t) \to f) && = 
{\cal N}_f \, | A_f |^2 \, \frac{1 + \lt| \lambda_f \rt|^2}{2}
        \, e^{-\Gamma_d t} \times   \nn \\
&& \lt[ \cosh \frac{\dg \, t}{2} \, + \,  
   \adi \, \cos ( \dm t )  
  + \adg \, \sinh \frac{\dg \, t}{2} 
  + \ami \, \sin \lt( \dm t \rt) \rt] ,   
\label{bd-f} \\
\!\!\!  \Gamma(\bar{B}_d(t) \to f) && =  
{\cal N}_f \, | \bar{A}_f |^2 \,  
        \frac{1 + \lt| \lambda_f \rt|^2}{2}\, 
  e^{-\Gamma_d t} \times  \nn \\
&& \lt[  
    \cosh \frac{\dg \, t}{2}  
  - \adi \, \cos ( \dm  t ) 
        + \adg \, \sinh \frac{\dg \, t}{2} 
        - \ami \, \sin ( \dm t ) \rt]  .
   \label{bdbar-f}
\earr
where the CP asymmetries are defined as
\begin{eqnarray}
\adi = \frac{1- \lt| \lambda_f \rt|^2}{1+ \lt| \lambda_f 
        \rt|^2} , && \qquad
\ami = - \frac{2\, {\rm Im} \lambda_f}{1+ \lt| \lambda_f \rt|^2} 
   \quad \mbox{and} \quad
\adg = - \frac{2\, {\rm Re} \lambda_f}{1+ \lt| \lambda_f   \rt|^2} 
        \, , 
\label{3acp} 
\end{eqnarray}
and ${\cal N}_f$ is a time-independent normalization factor.

In the case of semileptonic decays, $f \equiv \{D \ell^+ \nu \}$,
so that $\bar{A}_f = 0$ and hence $\lambda_f=0$. The time evolution
(\ref{bd-f}) then becomes
\barr
\Gamma(B_d(t) \to f) & \propto & e^{-\Gamma_d t}
\lt[ \cosh \frac{\dg \, t}{2} \, + \,  \cos ( \dm \, t ) \rt]~, \\
& \propto &   e^{-\gl t} + e^{-\gh t} +
\mbox{  oscillating terms }~, 
\label{bd-sl}
\earr
so that for semileptonic decays, we have $b_{SL}=0$. 
Note that
$b=0$ is true for all self-tagging modes, so that all the arguments
below for semileptonic modes hold true also for all the self-tagging
decay modes.

For the decays to CP eigenstates that proceed only through tree
processes (and have zero or negligible penguin contribution), 
we have $\lambda_f = \pm  (1 - \delta^d) e^{-2 i \beta} $ 
(the two signs ``$+$''  and ``$-$''
correspond to CP-even and CP-odd final states respectively).
Then (\ref{bd-f}) gives
\barr
\Gamma(B_d(t) \to f) & \propto & e^{-\Gamma_d t} 
\lt[ \cosh \frac{\dg \, t}{2} \,  
  \mp \cos(2 \beta) \, \sinh \frac{\dg \, t}{2} 
  \pm \sin (2 \beta) \, \sin \lt( \dm \, t \rt) \rt]~ , \\
& \propto &
e^{-\gl t} (1 \pm \cos(2 \beta)) + 
e^{-\gh t} (1 \mp \cos(2 \beta)) + 
\mbox{  oscillating terms }~, 
\label{bd-cp}
\earr
where we have neglected the small corrections due to $\delta^d$.
Thus, for CP eigenstates, we have 
$b_{CP+} = +\cos(2\beta)$ and 
$b_{CP-} = -\cos(2\beta)$. 

The ratio between the two lifetimes 
$\tau_{CP\pm}$ and $\tau_{SL}$ is then
\beq
\frac{\tau_{SL}}{\tau_{CP\pm}} = 1 \pm \frac{\cos(2\beta)}{2} 
\frac{\dg}{\Gamma_d}
+ {\cal O} \left[ (\dg/\Gamma_d)^2 \right]
~~.
\label{cp-sl}
\eeq
The measurement of these two lifetimes 
should be able to give us a value of $|\dg|$,
since $|\cos(2\beta)|$ will already be known to a good accuracy
by that time.

Note that it is also possible to measure the ratio of the lifetimes
$\tau_{CP-}$ and $\tau_{CP+}$: 
\beq
\frac{\tau_{CP-}}{\tau_{CP+}} = 1 +\cos (2\beta) \frac{\dg}{\Gamma_d}
+ {\cal O} \left[ (\dg/\Gamma_d)^2 \right]~~.
\label{cp-+1}
\eeq
Although the deviation of the ratio from 1.0 
in this case is larger by a factor of 2, using the effective
semileptonic lifetime instead of one of the CP eigenstates 
would still be the favoured method. 
This is because the CP specific decay modes of $B_d$ 
(e.g. $J/\psi K_{S(L)}, D^+ D^-$) have smaller
branching ratios than the semileptonic modes.
In addition, the ``semileptonic'' data sample may be enhanced 
by including the self-tagging decay modes (e.g. $D_s^{(*)+}D^{(*)-}$)
that also have large branching ratios. 
After 5 years of LHC, 
we should have about $5 \times 10^5$ events of 
$J/\psi K_S$, whereas the number of semileptonic decays  
at LHCb alone that will be directly useful in the lifetime 
measurements is expected to be more than $10^6$ per year, 
even with conservative estimates of efficiencies.

\subsection{Transversity angle distribution in 
$B_d \to J/\psi K^*$ }
\label{jpsik*}

The decays $B_d \to VV$ (where $VV$ is a flavour-blind final state
consisting of two vector mesons) take place both
through CP-even and CP-odd channels. 
Since the angular information is available here in addition to
the time information, these decay modes are not subject to the
constraints of the theorem in Sec.~\ref{quad}, and
quantities sensitive linearly to $\dg/\Gamma_d$ can be 
obtained through a single final state.
This cancels out many systematic uncertainties, and hence
these modes can be extremely useful 
as long as the direct CP violation is negligible, and
we can disentangle the CP-even and CP-odd final
states from each other. This separation 
can indeed be achieved through the
transversity angle distribution (\cite{ddlr}--\cite{durham}).

We illustrate the procedure with the example of 
$B_d \to J/\psi(\ell^+ \ell^-) K^*(K_S \pi^0)$.
The most general amplitude for the decay  $B \to J/\psi K^*$
is given in terms of the polarizations $\epsilon_{J/\psi},
\epsilon_{K^\ast}$ of the two vector mesons: 
\beq
A(B_d \to J/\psi K^\ast) =  A_0 
\left(\frac{m_{K^\ast}}{E_{K^\ast}} \right) 
\epsilon^{*L}_{J/\psi} \epsilon^{*L}_{K^\ast} - 
\frac{A_\parallel}{\sqrt{2}}~ 
\epsilon^{*T}_{J/\psi} \cdot \epsilon^{*T}_{K^\ast}
- i \frac{A_\perp}{\sqrt{2}}~
\epsilon^*_{J/\psi} \times \epsilon^*_{K^\ast} 
\cdot {\bf \hat p}~~,
\label{general}
\eeq
where $E_{K^\ast}$ is the energy of the $K^\ast$ 
and ${\bf \hat p}$ the unit vector in the direction of 
$K^\ast$ in the $J/\psi$ rest frame.
The superscripts $L$ and $T$ represent the longitudinal
and transverse components respectively.
Since the direct CP violation in this mode is negligible,  
the amplitudes $A_0$ and $A_\parallel$ are CP-even,
whereas $A_\perp$ is CP-odd.
Let us define the angles as follows.
Let the $x$ axis be the direction of $K^\ast$ in the $J/\psi$ 
rest frame, and the $z$ axis be perpendicular to the decay plane of 
$K^\ast \to K_S \pi^0$, with the positive $y$ direction
chosen such that $p_y(K_S) \geq 0$.
Then we define $(\theta, \varphi)$ as the decay direction of 
$\ell^+$ in the $J/ \psi$ rest frame and $\psi$ as the angle made by
$K_S$ with the $x$ axis in the $K^\ast$ rest frame.

Here $\theta$ is the transversity angle, {\it i.e.} the angular
distribution in $\theta$ can separate CP-even and CP-odd
components of the final state. The angular distribution is given by
\cite{ddf1}
\beq
\frac{d\Gamma[B_d \to J/\psi(\ell^+ \ell^-) K^*(K_S \pi^0)]}
{d\cos \theta}
= \frac{3}{8} |A_+(t)|^2 (1 + \cos^2 \theta) +
\frac{3}{4} |A_-(t)|^2 \sin^2 \theta
\label{one-angle}
\eeq
where $|A_+(t)|^2 \equiv |A_0(t)|^2 + |A_\parallel(t)|^2$ is the
CP-even component and $|A_-(t)|^2 \equiv |A_\perp(t)|^2$ 
the CP-odd one. These two components can be separated 
from the angular distribution (\ref{one-angle}) through a 
likelihood fit or through the method of angular moments 
\cite{ddf1,sen}\footnote{In~\cite{durham} we suggested to use
the CP-odd--CP-even interference in the decay $B \to J/\psi K^*$ to 
measure the value of $\dg/\Gamma_d$.
However, it involves tagged measurements in addition to two- or 
three-angle distributions, and hence is not as attractive as the 
untagged measurements described here.}.

The time evolutions of the CP-even and CP-odd components are given by
\barr
|A_+(t)|^2 & = & |A_+(0)|^2 \left[ \cos^2 \beta~ e^{-\gl t} 
+ \sin^2 \beta ~e^{-\gh t} +
e^{-\Gamma_d t}\sin(\Delta M_d t) \sin(2\beta)\right]~~,
\label{aplussq} \\
|A_-(t)|^2 & = & |A_-(0)|^2 \left[
\sin^2 \beta ~ e^{-\gl t} + 
\cos^2 \beta ~e^{-\gh t} -
e^{-\Gamma_d t}\sin(\Delta M_d t)\sin(2\beta)\right]~~.
\label{aminussq} 
\earr
These are the same as the time evolutions in (\ref{bd-cp}). 
The difference in the untagged lifetimes of the two components,
\beq
\frac{\tau_{CP-}}{\tau_{CP+}} = 1 +\cos (2\beta) \frac{\dg}{\Gamma_d}
+ {\cal O} \left[ (\dg/\Gamma_d)^2 \right]~~,
\label{cp-+}
\eeq
is linear in the lifetime difference $\dg$.

The disentanglement of the CP-even and CP-odd components from the
angular distribution is a statistically efficient process
\cite{sen}. In fact,
in the $B_s$ system, the angular distribution of 
$B_s \to J/\psi(\ell^+ \ell^-) \phi (K^+ K^-)$ can be used for
determining the lifetime difference $\Delta \Gamma_s$, and is the
preferred mode for measuring this quantity.

The mode $J/\psi K^*$ suffers from the presence of a $\pi^0$ in the final
state, which may be missed by the detector, thus introducing a source of
systematic error that needs to be minimized.

\subsection{Untagged asymmetry between 
$B \to J/\psi K_S$ and $J/\psi K_L$}
\label{kl-ks}

Two of the decay modes of $B_d$ that have been well
explored experimentally
(because of their usefulness in measuring $\beta$) are
$B \to J/\psi K_S$ and $J/\psi K_L$. Here we 
show that the time-dependent asymmetry between the
decay rates of these modes is a quantity linear in
$\dg/\Gamma_d$, and therefore within the domain of experimental
feasibility.

Let us define
$$
A(B_d\rightarrow J/\psi K_S) = A_S ,~~~ 
A(\bar{B}_d\rightarrow J/\psi K_S) = \bar{A}_S,$$
$$
A(B_d\rightarrow J/\psi K_L) = A_L ,~~~ 
A(\bar{B}_d\rightarrow J/\psi K_L) = \bar{A}_L,
$$
so that using
\beq
|K_S \rangle  = (1+ \epsilon) |K^0 \rangle  + 
(1-\epsilon) | \bar{K}^0 \rangle  ~, ~~
|K_L \rangle  = (1+ \epsilon) |K^0 \rangle  -
(1-\epsilon) | \bar{K}^0 \rangle  ~, ~
\label{kskl}
\eeq
we can write (with the phase convention ${\rm Arg}(q/p)=0$)
\barr
A_S & = A_L = & A e^{i\beta} (1 + a_p e^{i\theta} 
e^{i \Delta \gamma}) (1 + \epsilon) ~~, \nn \\
\bar{A}_S & = -\bar{A}_L = & A e^{-i\beta} (1 + a_p e^{i\theta} 
e^{-i \Delta \gamma}) (1 - \epsilon) ~~,
\label{asal}
\earr
where $a_p e^{i\theta} e^{i \Delta \gamma}$ is the ratio of
contributions that involve the CKM factors
$V_{cb}^* V_{cs}$ and $V_{tb}^* V_{td}$ respectively.
The latter contribution (penguin) is highly suppressed
with respect to the former one (tree): the value of
$a_p$ is less than a percent.  
Here $\theta$ is the strong phase and 
$\Delta \gamma \equiv {\rm Arg}
(V_{tb}^* V_{ts}/V_{cb}^* V_{cd}) \approx -0.015$
in the SM. From (\ref{lambda-def}), (\ref{q/p}) and (\ref{asal}), 
we get
\beq
\lambda_S = - \lambda_L \approx  - (1-\delta^d) e^{-2 i \beta} 
(1 - 2 \epsilon - 2 i \sin \Delta \gamma ~ a_p e^{i \theta})
\equiv -e^{-2 i \beta} (1 - 2 \bar{\epsilon})~~,
\label{eff-eps}
\eeq
where $\bar{\epsilon}$ is an effective complex parameter that
absorbs all the small theoretical uncertainties.

When the production asymmetry between $B_d$ and $\bar{B}_d$ 
is zero (as is the case at the $B$ factories), the 
untagged rate of decay is
\barr
\Gamma[B_{un} \to J/\psi K_S(K_L)] & \approx &
{\cal N} |A_S|^2 (1-2 {\rm Re}(\bar{\epsilon})) e^{-\Gamma_d t} 
\times \nn \\
& & \lt[ \cosh \left(\frac{\dg t}{2} \right) + 
\aa_{\dg} \sinh \left(\frac{\dg t}{2} \right)
\rt].
\earr 
The only difference between the decay to $K_S$ and 
that to $K_L$
is the sign of $\aa_{\Delta \Gamma}$:
\beq
\aa_{\Delta \Gamma} (K_S) = -\aa_{\Delta \Gamma} (K_L) =
\cos(2 \beta) - 2~ {\rm Im} (\bar{\epsilon}) \sin(2 \beta) ~~.
\eeq
The untagged time-dependent asymmetry between 
$B_{un} \rightarrow J/\psi K_S $ and $K_L$ is
\barr
{\cal A}(K_L,K_S) & \equiv & 
\frac{\Gamma (B_{un}(t)\rightarrow J/\psi K_S ) 
- \Gamma (B_{un}(t)\rightarrow J/\psi K_L )} 
{\Gamma (B_{un}(t)\rightarrow J/\psi K_S ) 
+ \Gamma (B_{un}(t)\rightarrow J/\psi K_L )} \\
& = & \cos(2 \beta) \tanh \lt(\frac{\dg t}{2} \rt)
\lt [1 - 2~ {\rm Im} (\bar{\epsilon}) \tan(2 \beta) \rt] \\
& \approx &  \cos(2 \beta) \tanh \lt(\frac{\dg t}{2} \rt)~~.
\earr
Thus, the measurement of this asymmetry will enable us to
determine $|\dg|$, given sufficient statistics and a measurement
of $\sin 2 \beta$.

The factor limiting the accuracy of the above asymmetry is the
measurement of $\Gamma (B_{un}(t)\rightarrow J/\psi K_L)$.
At the B factories, $K_L$ may be detected through its hadronic
interactions in the calorimeter, and though its energy is poorly
measured, the corresponding detection of $J/\psi$ can help in 
reducing the background. However, the number of events available
may be too small for an accurate measurement.
In the hadronic machines this decay has a high background, 
so the systematic errors in the measurement may be too large for this 
method to be of practical use.

\subsection{Effect on the measurement of $\sin(2 \beta)$}
\label{gold}

The time-dependent CP asymmetry measured through the 
``gold-plated'' mode $B_d \to J/\psi K_S$ is
\cite{babar-beta,belle-beta}
\barr
\aa_{CP} & = &\frac{\Gamma[\bar{B}_d(t) \to J/\psi K_S] - 
\Gamma[B_d(t) \to J/\psi K_S]}
{\Gamma[\bar{B}_d(t) \to J/\psi K_S] +
\Gamma[B_d(t) \to J/\psi K_S]}
\label{acp-ks} \\
 & \approx &  \sin(\Delta m_d t) \sin(2\beta)~~,
\label{acp-approx}
\earr
which is valid when the lifetime difference, the direct CP
violation, and the mixing in the neutral $K$ mesons is
neglected. As the accuracy of this measurement increases,
the corrections due to these factors will need to be taken 
into account. Keeping only linear terms in the small quantities
$\bar{\epsilon}$ and $\Delta \Gamma_d$, we get
\barr
\aa_{CP} & = & \sin(\Delta m t) 
        \sin(2\beta) \left[ 1 - \sinh \left( \frac{\dg t}{2} \right)
        \cos(2 \beta) \right] 
\label{dg-corr} \\
& & + 2 {\rm Re}(\bar{\epsilon}) 
\left[ -1 + \sin^2(2 \beta) \sin^2(\Delta m t) - \cos(\Delta m t) 
\right]
\label{re-eps-corr} \\
& & + 2 {\rm Im}(\bar{\epsilon}) \cos(2\beta) \sin(\Delta m t) ~~.
\label{im-eps-corr} 
\earr
The first term in (\ref{dg-corr}) represents the  standard approximation 
used (\ref{acp-approx}) and the correction due to the 
lifetime difference $\dg$.
The rest of the terms [(\ref{re-eps-corr}) and (\ref{im-eps-corr})]
include corrections due to the CP violation in $B$--$\bar{B}$ and
$K$--$\bar{K}$ mixings, which are of the same order as
$\dg/Gamma_d$.

In the future experiments 
that aim to measure $\beta$ to an accuracy of 0.005
\cite{lhc}, the correction terms need to be taken into account.
The corrections due to $\bar{\epsilon}$ and 
$\dg$ will form a major part of the systematic error, which
can be taken care of by a simultaneous fit to
$\sin(2\beta), \dg$ and $\bar{\epsilon}$.
The BaBar collaboration gives the bound on the coefficient
of $\cos(\Delta m t)$ in (\ref{re-eps-corr}), while neglecting the other
correction terms \cite{babar-direct}. 
When the measurements are accurate enough to
measure the $\cos(\Delta m t)$ term, the rest of the terms would also 
have come within the  domain of measurability.
For a correct treatment at this level of accuracy, the complete
expression for ${\cal A}_{CP}$ above (\ref{dg-corr}--\ref{im-eps-corr})
needs to be used.

\subsection{Tagged measurements}
\label{tagged}

Until now, we have discussed only the untagged measurements.
Taking into account the oscillating part of the time evolution
of the decay rate, we have the decay rate in general as
\beq
g(t) = f(t) + C e^{-\Gamma_d t} \sin(\Delta m t + \Phi)~,
\eeq
where $f(t)$ is the untagged decay rate as defined in 
(\ref{ft}), $C$ a constant and $\Phi$ a phase.
The lifetime of the oscillating part is an additional lifetime
measurement,  
which opens up the possibility of being able to determine
$\dg/\Gamma_d$ through only one final state (and without 
angular distributions as in Sec.~\ref{jpsik*}).

In the case of the semileptonic decays, this strategy 
fails since the semileptonic width measured with the
untagged sample is
\beq
\Gamma_{SL}= \frac{(\Gamma_L+ \Gamma_H) \Gamma_L \Gamma_H }
{(\Gamma_L)^2 + (\Gamma_H)^2}
= \Gamma_d 
\frac{1 - \frac{1}{4} \left( \frac{\dg}{\Gamma_d} \right) ^2}
{1 + \frac{1}{4} \left( \frac{\dg}{\Gamma_d} \right) ^2}~~,
\label{gam-sl}
\eeq
so that 
\beq
\Gamma_{SL}/\Gamma_d = 1 +  
{\cal O} \left[ (\dg/\Gamma_d)^2 \right]~~.
\eeq
Thus the semileptonic decays would provide sensitivity 
only to quadratic terms in $\dg/\Gamma_d$, 
even if it were possible to use the tagged measurements
efficiently.

However, the untagged lifetime measured through the decay to
a CP eigenstate is
\beq
\tau_{CP\pm} \approx \frac{1}{\Gamma_d} 
\left( 1 \mp \frac{\cos(2\beta)}{2} \frac{\dg}{\Gamma_d}
\right)~~,
\eeq
so that it differs from the lifetime of the oscillating part
($\tau_d \equiv 1/\Gamma_d$) by terms linear in $\dg/\Gamma_d$.
Thus, the tagged measurements of a CP-even or CP-odd final 
state ($D^+ D^-$, $J/\psi K_S$, $J/\psi K_L$, etc.)
can measure $\dg/\Gamma_d$ by themselves.

The mistag fraction is the main limiting factor on the accuracy of this
measurement, and the tagging efficiency limits the number of
events available. It is indeed possible that the $\tau_d$
measurement through the semileptonic decays will be more accurate
than that through the oscillating part of the CP-specific final
state. This then reduces to the method suggested in 
Sec.~\ref{semilep}. For further experimental details on a tagged measurement 
of $\dg/\Gamma_d$ we refer the reader to reference \cite{Babarnote}.

\section{Lifetime differences in $B_s$ and $B_d$ systems}
\label{contrast}

The calculations of the lifetime difference in $B_d$ (as performed
here) and in the $B_s$ system (as in \cite{BBD,BBGLN}) run along
similar lines. However, there are some subtle differences involved,
due to the values of the different CKM elements involved, which
have significant consequences. In particular, whereas the upper
bound on the value of $\Delta\Gamma_s$ (including the effects of
new physics) is the value of $\Delta\Gamma_s({\rm SM})$ \cite{grossman},
such an upper bound on $\Delta\Gamma_d$ can be established only under
certain conditions and involves a multiplicative
factor in addition to $\Delta\Gamma_d({\rm SM})$. Also, whereas 
the difference in lifetimes of CP-specific final states in
the $B_s$ system cannot resolve the discrete ambiguity in
the $B_s$--$\bar{B}_s$ mixing phase, the corresponding measurement
in the $B_d$ system can resolve the discrete ambiguity in
the $B_d$--$\bar{B}_d$ mixing phase.
Let us elaborate on these two differences in this section.

\subsection{Upper bounds on $\Delta\Gamma_{d(s)}$
in the presence of new physics}
\label{upper-bd}

For convenience, let us define 
$\Theta_q \equiv {\rm Arg}(\Gamma_{21})_q , 
\Phi_q \equiv {\rm Arg}(M_{21})_q $,
where $q \in \{ d,s \}$. Then we can write 
\beq
\Delta\Gamma_q = - 2 |\Gamma_{21}|_q  \cos(\Theta_q - \Phi_q)~~.
\label{theta-phi}
\eeq
Since the contribution to $\Gamma_{21}$ comes only from tree
diagrams, we expect the effect of new physics on this quantity 
to be very small. We therefore take
$|\Gamma_{21}|_q$ and $\Theta_q$ to be unaffected by new physics. 
On the other hand, the mixing phase $\Phi_q$ appears from loop
diagrams and can therefore be very sensitive to new physics.

Let us first consider the $B_s$ system.
Here $\Gamma_{21}$ may be written in the form
\beq
\Gamma_{21}(B_s) = - {\cal N} [(V_{cb}^* V_{cs})^2 f(z,z) 
                + 2 (V_{cb}^* V_{cs})(V_{ub}^* V_{us}) f(z,0) 
                        + (V_{ub}^* V_{us})^2 f(0,0)]
\label{bs-g12}
\eeq
where ${\cal N}$ is a positive normalization constant and
$f(x,y)$ are the hadronic factors that do not depend on the
CKM matrix elements. In the limit $z \equiv m_c^2/m_b^2 \to 0$,
we get $f(z,z) = f(z,0) = f(0,0)$. 
Even with $z \neq 0$, we expect
all the $f$'s to have similar magnitudes.
On the other hand,
the CKM elements involved in (\ref{bs-g12}) obey the hierarchy
$(V_{cb}^* V_{cs})^2 \sim \lambda^4~, \ 
 (V_{cb}^* V_{cs})(V_{ub}^* V_{us}) \sim \lambda^6~, \ 
 (V_{ub}^* V_{us})^2 \sim \lambda^8$. 
The term involving $(V_{cb}^* V_{cs})^2$ then
dominates in (\ref{bs-g12}), and we can write
\beq
\Gamma_{21}(B_s) = -{\cal N}  (V_{cb}^* V_{cs})^2 f(z,z)
 [ 1 + O(\lambda^2) ]~.
\label{bs-g12-appr}
\eeq
Since the $f$'s are real positive functions,
we have $\Theta_s \approx \pi + {\rm Arg}(V_{cb}^* V_{cs})^2  $.
Then,
\beq
\Delta \Gamma_s = 2 |\Gamma_{21}|_s 
\cos[ {\rm Arg}(V_{cb}^* V_{cs})^2 - \Phi_s]~~.
\label{dgs}
\eeq
In SM, $\Phi_s = {\rm Arg}(V_{tb}^* V_{ts})^2$, therefore the argument
of the cosine term in (\ref{dgs}) is given by
${\rm Arg}[(V_{cb}^* V_{cs})^2/(V_{tb}^* V_{ts})^2] = - 2\Delta\gamma
\approx 0.03$. Thus in SM, we have
\beq
\Delta \Gamma_s({\rm SM}) = 2 |\Gamma_{21}|_s \cos(2\Delta\gamma)~~.
\label{dgs-sm}
\eeq
The effect of new physics on $\Delta\Gamma_s$ can then be bounded
by giving an upper bound on $\Delta\Gamma_s$:
\beq
\Delta\Gamma_s \leq \frac{\Delta \Gamma_s({\rm SM})}{\cos(2\Delta\gamma)}
\approx \Delta \Gamma_s({\rm SM})~~.
\label{dgs-bd}
\eeq
Thus, the value of $\Delta \Gamma_s$ can only decrease in the
presence of new physics \cite{grossman}.

In the case of the $B_d$ system, the situation is slightly different.
As in the $B_s$ case, we can write
\beq
\Gamma_{21}(B_d) = -{\cal N} [(V_{cb}^* V_{cd})^2 f(z,z) 
                + 2 (V_{cb}^* V_{cd})(V_{ub}^* V_{ud}) f(z,0) 
                        + (V_{ub}^* V_{ud})^2 f(0,0)]
\label{bd-g12-uc}
\eeq
where the normalizing factor ${\cal N}$ and the hadronic
factors $f$ are the same as in the $B_s$ case in the
limit of the U-spin symmetry (see \cite{uspin}) 
and therefore have similar magnitudes.
The CKM elements involved in (\ref{bd-g12-uc}) do not obey a hierarchy
similar to the $B_s$ case: instead we have
$(V_{cb}^* V_{cd})^2 \sim (V_{cb}^* V_{cd})(V_{ub}^* V_{ud})
\sim (V_{ub}^* V_{ud})^2 \sim \lambda^6$. 
Then no single term
in (\ref{bd-g12-uc}) can dominate. We can, however, use the unitarity
of the CKM matrix\footnote{ 
We note that this assumption of the unitarity for a 
three-generation CKM matrix is quite general, because most popular 
new physics models, including supersymmetric models,  preserve 
the three-generation CKM unitarity.
The present CKM values, constrained from various experiments,         
are completely consistent with the unitarity for 
the three-generation CKM matrix.
Moreover, one can show that the non-unitary effects within 
the three-generation
CKM, which can stem from the fourth generation or E(6)-inspired models with   
one singlet down-type quark, are $\lsim \lambda^4$, once we assume a
Wolfenstein-type hierarchical structure for the extended CKM matrix.}
to rearrange (\ref{bd-g12-uc}) in the form
\bea
\Gamma_{21}(B_d) & = &  -{\cal N}  
\Bigl[ (V_{cb}^* V_{cd})^2 [f(z,z)-2f(z,0)+f(0,0)]  \nn \\ 
 & &  + 2 (V_{cb}^* V_{cd})(V_{tb}^* V_{td}) [ f(0,0) -f(z,0)] 
                        + (V_{tb}^* V_{td})^2 f(0,0) \Bigr]~~.
\label{bd-g12}
\eea
Note that in the limit of $z\to 0$, all the factors $f$ are 
identical and hence the coefficients of $(V_{cb}^* V_{cd})^2$
and $(V_{cb}^* V_{cd})(V_{tb}^* V_{td})$ vanish. The last term 
in (\ref{bd-g12}) is 
then left over as the dominating one, and we get
\beq
\Gamma_{21}(B_d) \approx -{\cal N} (V_{tb}^* V_{td})^2 f(0,0)~,
\label{g12-tt}
\eeq
thus giving $\Theta_d \approx \pi + {\rm Arg}(V_{tb}^* V_{td})^2$.
However, the finite value of $z \approx 0.1$ may give large corrections 
to this value.


We have already shown that (\ref{g12-tt}) may be written in the forms
(\ref{ct-form}) or (\ref{ut-form}).
From this, we get
$\Theta_d = \pi + {\rm Arg}(V_{tb}^* V_{td})^2 + {\rm Arg}(1 + \delta f_i)$,
where $i \in \{u,c\}$.
Using (\ref{theta-phi}), we then have
\beq
\Delta \Gamma_d  \approx  2 |\Gamma_{21}|_d 
\cos[ {\rm Arg}(V_{tb}^* V_{td})^2 - \Phi_d + {\rm Arg}(1 + \delta f_i)]~~.
\label{dgd}
\eeq
In SM, $\Phi_d = {\rm Arg} (V_{tb}^* V_{td})^2$, so that
\beq
\Delta \Gamma_d ({\rm SM}) \approx  2 |\Gamma_{21}|_d 
\cos[ {\rm Arg}(1 + \delta f_i)]~~,
\label{dgd-sm}
\eeq  
and an upper bound for $\dg$ can be written in terms of $\dg$(SM) as
\beq
\Delta \Gamma_d \leq 2 |\Gamma_{21}|_d =  \frac{\Delta \Gamma_d(SM)}
{\cos[{\rm Arg}(1 + \delta f_i)]}~~.
\label{dgd-bd}
\eeq
We can calculate the
bound (\ref{dgd-bd}) in terms of the extent of the higher order NLO
corrections. 
Estimating $|{\cal F}_{ct}| < 0.4$ (corresponding to the error analysis 
in Sec.~\ref{numerical}), we get $|{\rm Arg}(1 + \delta f_u)| < 0.6$,
so that we have the bound 
$\Delta \Gamma_d < 1.2 ~ \Delta \Gamma_d(SM)$. 
Note that this bound is valid only in the range of ${\cal F}_{ct}$
estimated above. A complete NLO calculation will be able to give a
stronger bound.

Thus in the case of the $B_d$ system, we have an upper bound
analogous to the one in the $B_s$ system only under certain conditions.
Moreover, the reasons behind
the existence of these two upper bounds differ. 
Whereas in the $B_s$ case it follows directly from
the hierarchy in the CKM elements, in the $B_d$ case
it depends on the values of the hadronic terms.
Note that whereas unitarity was not needed in the $B_s$ case,
the assumption that $(\Gamma_{21})_q$ is unaffected by new physics
is required in both the cases.

\subsection{Mixing phase: new physics and discrete ambiguity}
\label{discrete}

The $B_d$--$\bar{B}_d$ mixing phase $\Phi_d$ is efficiently
measured through the decay modes $J/\psi K_S$ and $J/\psi K_L$. 
If we take the new physics effects into account, the time-dependent
asymmetry (\ref{acp-ks}) is 
${\cal A}_{CP} = - \sin(\Delta M_d t) \sin(\Phi_d)$,
which reduces to (\ref{acp-approx}) in the SM, where $\Phi_d = -2\beta$. 
The measurement of $\sin(\Phi_d)$ still allows for a discrete ambiguity
$\Phi_d \leftrightarrow \pi - \Phi_d$. 
Whenever a discrete ambiguity
in $\beta$ is referred to ($\beta \leftrightarrow \pi/2 -\beta$)
in this paper (or in the literature), strictly speaking we are 
talking about the discrete ambiguity 
$\Phi_d \leftrightarrow \pi - \Phi_d$. In this section, we
shall use the notation $\Phi_d$ instead of $2\beta$ in order to
illustrate the comparison with the corresponding quantities
in the $B_s$ system.

Getting rid of the above discrete ambiguity is a way of
uncovering a possible signal of new physics\footnote{
In SM, the value of $\Phi_d$ must match with the phase of
the $b \to d$ penguin. However, the direct measurement of the
latter phase is not theoretically clean \cite{kly}, so the
preferred way is to compare the measured value of $\Phi_d$ with
the value of $2\beta$ determined through an ``unitarity 
fit'' for all the CKM parameters \cite{ckm-fit}.}.
Ways to get rid of this ambiguity have been suggested in literature, 
using the comparison of CP asymmetries in $J/\psi K_S$ and $\pi \pi$ 
\cite{gnw}, time dependent CP asymmetries in $B_s \to \rho K_S$ 
\cite{gro-qui} and in $B_s \to \pi K, ~K K$ \cite{kly1}, 
angular distributions and U-spin symmetry arguments 
\cite{ddf2}, or cascade decays $B \to D \to K$ \cite{kayser}.  
The measurement of $\Phi_d$ through the measurements involving
$\dg$ is unique in the sense that it uses only untagged
measurements.

In Sections \ref{semilep} and \ref{jpsik*}, we have seen that the ratio of 
two effective lifetimes can enable us to measure the quantity
$\Delta\Gamma_{obs(d)} \equiv \cos(2\beta) \Delta\Gamma_d/\Gamma$.
In the presence of new physics, this quantity is in fact
(see eq.~\ref{theta-phi})
\beq
\Delta\Gamma_{obs(d)} =  - 2 (|\Gamma_{21}|_d/\Gamma_d)  
\cos(\Phi_d) \cos(\Theta_d - \Phi_d)~~.
\label{dg-obs-d}
\eeq
In SM, we get 
\begin{equation}
\Delta\Gamma_{obs(d)}({\rm SM}) = 2 (|\Gamma_{21}|_d/\Gamma_d) 
\cos(2 \beta) \cos[ {\rm Arg}(1 + \delta f_i) ]~~.
\end{equation}
If $|\delta f_i| < 1.0$, we have $\cos[{\rm Arg}(1+\delta f)] > 0$ 
(in fact, from the fit in \cite{ckm-fit} and our error estimates, 
we have $\cos[{\rm Arg}(1+\delta f_u)] > 0.8$). 
Then $\Delta\Gamma_{obs(d)}({\rm SM})$ is predicted to be positive.
New physics is not expected to affect $\Theta_d$, but it may affect 
$\Phi_d$ in such a way as to make the combination 
$\cos(\Phi_d) \cos(\Theta_d - \Phi_d)$ change sign.
A negative sign of $\Delta\Gamma_{obs(d)}$ is a clear signal of
such new physics.

In addition, if $\Theta_d$ can be determined independently of 
the mixing in the $B_d$ system\footnote{For example,
if we assume that new physics does not affect the mixing in the 
neutral $K$ system, 
the fit to $|V_{ub}/V_{cb}|$ and $\epsilon_K$ 
(a part of the unitarity fit) determines 
${\rm Arg}(V_{tb}^* V_{td})$ independent of any mixing in the 
$B$ system. This gives us a measurement of 
$\Theta_d = \pi+ {\rm Arg}(V_{tb}^* V_{td})^2 + Arg(1 + \delta f_i)$
where $Arg(1 + \delta f_i)$ will be known to a good accuracy with a complete
NLO calculation.},
then measuring $\Delta\Gamma_{obs(d)}$
and using (\ref{dg-obs-d}) gives us two solutions for $\Phi_d$
in general: $\Phi_{1d}$ and $\Phi_{2d}$ such that
$\tan(\Phi_{1d}+ \Phi_{2d}) = \tan(\Theta_d)$.
As long as $\tan(\Theta_d) \neq 0$ (as the
unitarity fit suggests), at most one of 
the solutions will correspond to the value of 
$\sin(\Phi_{d})_{J/\psi K_S}$
obtained through ${\cal A}_{CP}(J/\psi K_S)$.
If none of $\Phi_{d1}$ or $\Phi_{d2}$ corresponds to
the value of $\sin(\Phi_{d})_{J/\psi K_S}$, there definitely 
is new physics in $B_d$ mixing \footnote{
Or any assumption that goes into the determination of
$\Theta_d$ is in question.}.
If one of $\Phi_{1d}$ or $\Phi_{2d}$  matches with 
$\sin(\Phi_{d})_{J/\psi K_S}$,
it gives the actual value of $\Phi_d$, and thus 
resolves the discrete ambiguity in principle.
In practice, this requires the knowledge of
$\Gamma_{21(d)}$ theoretically to a high precision
and having to measure $\Delta\Gamma_{obs(d)}$ to sufficient accuracy
to be able to distinguish between $\Phi_{1d}$ and $\Phi_{2d}$.
A complete NLO calculation is needed for the former. The latter may
be achieved at the LHC using the effective lifetimes of decays to
semileptonic final states and to $J/\psi K_S$.

Let us contrast this case with that in the $B_s$ system and show
 these  features above are  unique to the $B_d$ system: The
corresponding time-dependent asymmetry in the $B_s$ system 
is measured through the modes
$J/\psi \phi$ or $J/\psi \eta^{(')}$, 
which give the value of $\sin(\Phi_s)$, and 
therefore leave the discrete ambiguity
$\Phi_s \leftrightarrow \pi - \Phi_s$ unresolved. 
The ratio of 
two effective lifetimes in the $B_s$ system can enable us to 
measure the quantity
\barr
\Delta\Gamma_{obs(s)} & \equiv & \cos(\Phi_s)\Delta\Gamma_s/\Gamma \nn \\
& = & - 2 |\Gamma_{21}|_s/\Gamma_s  \cos(\Phi_s) \cos(\Theta_s - \Phi_s)
~~.
\earr
Since $\Theta_s \approx \pi + {\rm Arg}(V_{cb}^* V_{cs})^2 \approx \pi$,
we have
\beq
\Delta\Gamma_{obs(s)} \approx  2 |\Gamma_{21}|_s/\Gamma_s \cos^2(\Phi_s)~.
\eeq
This measurement thus still has the same discrete ambiguity
$\Phi_s \leftrightarrow \pi - \Phi_s$ as in the $J/\psi \phi$ 
(or $J/\psi \eta^{(')})$ case,
and the discrete ambiguity in the $B_s$ system is not resolved.

\section{Summary and conclusions}
\label{concl}

It has been known for many years that the $B_d$ system is 
a particularly good place to test the standard model explanation 
of CP violation through the unitary CKM matrix. 
The phase $2 \beta$ involved in the $B_d - \bar{B}_d$ mixing
is large, and hence the CP violation is expected to be larger
in the $B_d$ system in general, as compared to the $K$ or the
$B_s$ system. 
This feature has already been exploited in various methods for
extracting $\alpha,~\beta$ and $\gamma$, the angles of the 
unitarity triangle, by measuring CP-violating rate asymmetries 
in the decays of neutral $B_d$ mesons to a variety of final states.
In particular, the precise measurement of $\sin(2\beta)$ from  
the theoretically clean decay modes $B_d(t) \to J/\psi K_S(K_L)$ 
is a test of the SM, as well as the opportunity to search
for the presence of physics beyond the standard model.

The two mass eigenstates of the neutral $B_d$ system --- $B_H$ and
$B_L$ --- have slightly different lifetimes: the lifetime difference
is less than a percent. At the present accuracy of measurements,
this lifetime difference $\dg$ can well be ignored. As a result,
the measurement and the phenomenology of $\dg$ has been neglected
so far, as compared to the lifetime difference in the $B_s$ system
for example. However,
with the possibility of experiments with high time resolution and 
high statistics, such as the electronic asymmetric $B$ factories of 
BaBar, BELLE, and hadronic $B$ factories of CDF, LHC and BTeV, 
this quantity starts becoming more and more relevant.

Taking the effect of $\dg$ into account is important in two
aspects. On one hand, it affects the accurate measurements of 
crucial quantities like the CKM phase $\beta$ and therefore must
be measured in order to estimate and correct the error due to it.
On the other hand, the measurement of $\dg$ can lead to clear signal
for new physics.

Thus in addition to being the measurement of a well-defined
physical quantity which can be compared with the theoretical
prediction, the value of $\dg$ is important  for getting 
a firm grip on our understanding of CP violation.
It is therefore worthwhile to have a look at this quantity 
and make a realistic estimation of the  possibility of its 
measurement, as we do in this paper.

We estimate $\dg/\Gamma_d$ including $1/m_b$ contributions
and part of the next-to-leading order QCD corrections.
We find that adding the latter corrections decreases 
the value of $\dg/\Gamma_d$ computed at the leading order by almost
a factor of two. We get the final result as
$\Delta \Gamma_d / \Gamma_d = (2.6 ^{+1.2}_{-1.6}) \times 10^{-3}$,
or using another expansion of the NLO QCD corrections
$\Delta \Gamma_d / \Gamma_d = (3.0 ^{+0.9}_{-1.4}) \times 10^{-3}$,
where for the central value we have used the preliminary values 
for the bag factors from the JLQCD collaboration.
In the error estimation, we take into account the errors from
the uncertainties in the values of  
the CKM parameters, the bag parameters, the mass of the $b$ quark,
and the measured value of $x_d$.
The major sources of error are the scale dependence, the breaking
of the quark-hadron duality,  
and the error due to missing  terms in the NLO contribution.
We show that a complete NLO calculation is desirable.

The most obvious way of trying to measure the lifetime difference is 
through the semileptonic decays, however it runs into major
difficulties.
If only the non-oscillating (untagged) part of the time evolution of 
the decay is considered, we indeed have a combination of two exponential 
decays with different lifetimes. However, as we show in this paper, 
there is no observable quantity here that is linear in $\dg/\Gamma_d$.
The time measurements allow us to determine the quantity
$\tau_{SL} \equiv (1/\Gamma_d) [1 + {\cal O}(\dg/\Gamma_d)^2]$.
This decay mode is thus sensitive only to quantities quandratic in
$\dg/\Gamma_d$. So this method would involve measuring a quantity 
as small as $(\dg/\Gamma_d)^2 \sim 10^{-5}$, which is not practical.
The lifetime of the oscillating part is also $1/\Gamma_d$, so
adding the information from the oscillating part of the time evolution
does not help at all. This problem arises for all self-tagging decays.
Therefore, though self-tagging decays of $B_d$ have significant
branching ratios, they cannot by themselves be expected to give 
a measurement of $\dg/\Gamma_d$.

The time evolutions of $B_d$ decaying into CP eigenstates 
also involve both the lifetimes, since the $B_d - \bar{B}_d$ 
mixing phase ($2\beta$) is large, which implies that the CP eigenstates 
are appreciably different from the lifetime eigenstates. 
As a result, it is not possible to find a final state
to which the decay of $B_d$ involves only one of the decay widths
$\gl$ and $\gh$.
The non-oscillating part of the time evolution of decays to
CP eigenstates gives a quantity
$\tau_{CP_{\pm}} \equiv (1/\Gamma_d) [1 \pm (\cos(2\beta)/2) 
\dg/\Gamma_d +{\cal O}(\dg/\Gamma_d)^2]$, but the quantities
$\Gamma_d$ and $\dg$ cannot be separately determined
through this measurement,
and sensitivity to $(\dg/\Gamma_d)^2$ is necessary.
(Indeed, we explicitly prove a general theorem that shows that, for 
isotropic decays of $B_d$ to any final state, the untagged
measurements can only be sensitive to 
$(\dg/\Gamma_d)^2$.)

The oscillating part of the time evolution to CP eigenstates 
has a lifetime $1/\Gamma_d$ 
(to an accuracy of ${\cal O}(\dg/\Gamma_d)^2$).
Therefore, if this lifetime is measured accurately, it can be
combined with the measurement of $\tau_{CP_{\pm}}$ through the
untagged part to get a measurement linear in $\dg/\Gamma_d$.
However, the need for tagging, and consequent mistagging errors, 
reduce the efficiency of this method.

A viable option, perhaps the most efficient among the ones 
considered here, is to compare the measurements of the
untagged lifetimes  $\tau_{SL}$ and $\tau_{CP_{\pm}}$.
Since $\tau_{SL}$ is in fact the lifetime for all self-tagging
decays, and the branching ratios for self-tagging decays of
$B_d$ are much larger than the decays to CP eigenstates,
we expect that the most useful combination will be the
measurement of $\tau_{SL}$ through self-tagging decays and
that of $\tau_{CP_+}$ through $B_d \to J/\psi K_S$.

The untagged asymmetry between $B_d \to J/\psi K_S$ and
$B_d \to J/\psi K_L$ is a particular case of using the
combination of measurements of $\tau_{CP_+}$ and  $\tau_{CP_-}$,
which we analyze in detail. The effects of CP violation
in the mixing and decay of $B_d$, as well as the 
indirect CP violation in the $K$ system has been taken into
account. 
The usefulness of this method is limited by the uncertainties
in the measurement of $\Gamma(B_d \to J/\psi K_L$).


Since the theorem referred to above --- about a single untagged decay
being sensitive only to $(\dg/\Gamma_d)^2$ --- applies only to 
isotropic decays, decays of the type $B \to VV$ can still be used
by themselves to determine quantities linear in $\dg/\Gamma_d$.
A promising example is 
$B_d \to J/\psi (\to \ell^+ \ell^-)~K^*(\to K_s \pi^0)$.
The CP-odd and CP-even components in the final state can be 
disentangled through the transversity angle distribution, and both
$\tau_{CP_+}$ and  $\tau_{CP_-}$ can be determined through the
same decay. Since there is only one final state, many systematic
errors are reduced. The only undesirable feature of this decay mode 
is the presence of $\pi^0$ in the final state, which may be missed,
especially in the hadronic machines. 
The three angle distribution of the same decay mode can also be used to
obtain $\dg/\Gamma_d$ through the interference between CP-even and 
CP-odd final states. The three angle method is 
however not as efficient as the single angle distribution, since
one has to use tagged decays and more number of parameters need
to be fitted.

We also point out the interlinked nature of the accurate
measurements of $\beta$ and $\dg/\Gamma_d$ through the
conventional gold-plated decay. 
In the future experiments that aim to measure $\beta$ to an 
accuracy of 0.005 or better, the corrections due to
$\dg$ will form the major part of the systematic error, which
can be taken care of by a simultaneous fit to
$\sin(2\beta), \dg$ and 
an effective parameter $\bar{\epsilon}$ that comes from
a combination of CP violation in mixing in the $B_d$ and $K$ system.

All the combinations of untagged decay modes discussed here involve
measuring the quantity 
$\Delta \Gamma_{obs(d)} \propto (\cos(2\beta)/2) \dg/\Gamma_d$, wherein 
the value of $\dg/\Gamma_d$ also depends on $\beta$. 
The complete
dependence on $\beta$ is of the form 
$\cos(2\beta) \cos(\Theta_d + 2 \beta)$.
The sign of this quantity is positive in SM, but may be changed in the
presence of new physics. There is thus a potential for detecting physics
beyond the SM. 
Moreover, if 
the value of $\Theta_d$ is known independently of $B$ mixing, then since 
$\Delta \Gamma_{obs(d)}$ is not invariant under 
$\beta \leftrightarrow \pi/2 -\beta$, the discrete 
ambiguity in $\beta$ is resolved in principle.
Note that this feature is unique to
the $B_d$ system --- in the $B_s$ system for example, 
$\Delta \Gamma_{obs(s)}$ does not help in resolving the
corresponding discrete ambiguity in the 
$B_s$--$\bar{B}_s$ mixing phase.

It is known that, if $(\Gamma_{21})_s$ is unaffected by new
physics, then the value of $\Delta \Gamma_s$ in the $B_s$ 
system is bounded from above by its value as calculated in the
SM. In the $B_d$ system, this statement does not strictly hold true. 
However, if  $(\Gamma_{21})_d$ is unaffected by new physics and
the unitarity of the $3\times 3$ CKM matrix holds, then
an upper bound on the value of $\dg$ may be found. In the absence of
a complete NLO calculation, however, this bound is a weak one.

With the high statistics and accurate time resolution of the upcoming
experiments, the measurement of $\dg$ seems to be in the domain
of measurability. And given the rich phenomenology that comes with
it, it is certainly a worthwhile endeavor.

\vspace{-0.2cm}
\section*{Acknowledgments}

We would like to thank G. Buchalla for many useful  discussions
and for a reading of the manuscript.  
We also thank U. Nierste for critical comments, and 
C. Shepherd-Thermistocleous, 
Y. Wah and  H. Yamamoto for useful discussions concerning the 
experimental aspects. A.D. would like to thank the CERN Theory Division, where a large
fraction of this work was completed.
The work of C.S.K. was supported 
in part by  CHEP-SRC Program, Grant No. 20015-111-02-2
and Grant No. R03-2001-00010 of the KOSEF,
in part by BK21 Program and Grant No. 2001-042-D00022 of the KRF,
and in part by Yonsei Research Fund, Project No. 2001-1-0057.
 The work of T.Y. was supported 
in part by the US Department of Energy under Grant No. DE-FG02-97ER-41036.


\begin{thebibliography}{99}


\bibitem{aleksan}
R.~Aleksan, A.~Le Yaouanc, L.~Oliver, O.~P\'{e}ne and J.~C.~Raynal,
Phys.\ Lett.\ B {\bf 316} (1993) 567.


\bibitem{BBD}
M.~Beneke, G.~Buchalla and I.~Dunietz,
Phys.\ Rev.\ D {\bf 54} (1996) 4419
[hep-ph/9605259].


\bibitem{BBGLN}
M.~Beneke, G.~Buchalla, C.~Greub, A.~Lenz and U.~Nierste,
Phys.\ Lett.\ B {\bf 459} (1999) 631
[hep-ph/9808385].


\bibitem{isi95}I.~Dunietz,
Phys.\ Rev.\ D {\bf 52} (1995) 3048
[hep-ph/9501287].


\bibitem{dfn}
I.~Dunietz, R.~Fleischer and U.~Nierste,
hep-ph/0012219.


\bibitem{pdg} Particle Data Group,
D.E. Groom et al., Eur. Phys. J. {\bf C15} (2000) 1.

\bibitem{lhc}
P.~Ball {\it et al.},
hep-ph/0003238.

\bibitem{SM} 
R.~N.~Cahn and M.~P.~Worah,
Phys.\ Rev.\ D {\bf 60} (1999) 076006
[hep-ph/9904480].


\bibitem{SM2}
J.~S.~Hagelin,
Nucl.\ Phys.\ B {\bf 193} (1981) 123.
A.~J.~Buras, W.~Slominski and H.~Steger,
Nucl.\ Phys.\ B {\bf 245} (1984) 369.

\bibitem{NLLM12}
A.~J.~Buras, M.~Jamin and P.~H.~Weisz,
Nucl.\ Phys.\ B {\bf 347} (1990) 491.


\bibitem{WilsonNLL}
G. Altarelli, G. Curci, G. Martinelli and S. Petrarca,
Nucl. Phys. {\bf B187} (1981) 461;
A.J. Buras and P.H. Weisz, Nucl. Phys. {\bf B333} (1990) 66.

\bibitem{jlqcd}
S.~Hashimoto and N.~Yamada  [JLQCD collaboration],
hep-ph/0104080.


\bibitem{breport} 
Report of {\it Workshop on $B$ Physics at the 
Tevatron: Run II and Beyond} (unpublished). Available at \\ 
{\tt http://www-theory.lbl.gov/ $\tilde{ }$ ligeti/Brun2/report/drafts/draft.uu}, FERMILAB-Pub-01/197.




\bibitem{bbl} G.~Buchalla, A.~J.~Buras and M.~E.~Lautenbacher,
Rev.\ Mod.\ Phys.\  {\bf 68} (1996) 1125
[hep-ph/9512380].

\bibitem{ckm-fit}
S.~Mele,
hep-ph/0103040.


\bibitem{lenz} 
M.~Beneke and A.~Lenz,
J.\ Phys.\ G {\bf G27} (2001) 1219
[hep-ph/0012222].


\bibitem{ddlr}
A.~S.~Dighe, I.~Dunietz, H.~J.~Lipkin and J.~L.~Rosner,
Phys.\ Lett.\ B {\bf 369} (1996) 144
[hep-ph/9511363].

\bibitem{cskim}
C.~S. Kim, Y. G. Kim and C.-D. Lu, 
hep-ph/0102168; 
C.~S. Kim, Y. G. Kim, C.-D. Lu and T. Morozumi,
Phys.\ Rev.\ D {\bf 62} (2000) 034013
[hep-ph/0001151]. 

\bibitem{durham}
T.~Hurth {\it et al.},
J.\ Phys.\ G {\bf G27} (2001) 1277
[hep-ph/0102159].


\bibitem{ddf1}
A.~S.~Dighe, I.~Dunietz and R.~Fleischer,
Eur.\ Phys.\ J.\ C {\bf 6} (1999) 647
[hep-ph/9804253].

\bibitem{sen}
A.~Dighe and S.~Sen,
Phys.\ Rev.\ D {\bf 59} (1999) 074002
[hep-ph/9810381].



\bibitem{babar-beta}
B.~Aubert {\it et al.}  [BaBar Collaboration],
Phys.\ Rev.\ Lett.\  {\bf 86} (2001) 2515
[hep-ex/0102030].

\bibitem{belle-beta}
A.~Abashian {\it et al.}  [BELLE Collaboration],
Phys.\ Rev.\ Lett.\  {\bf 86} (2001) 2509
[hep-ex/0102018].


\bibitem{babar-direct} 
B.~Aubert {\it et al.}  [BABAR Collaboration],
Phys.\ Rev.\ Lett.\  {\bf 87} (2001) 091801


\bibitem{Babarnote}
S.~Petrak, {\it Reach of $\Delta \Gamma$ Measurements for $B_d^0$ and $B_s^0$},
BaBar Note 496 (unpublished). 



\bibitem{grossman}
Y.~Grossman,
Phys.\ Lett.\ B {\bf 380} (1996) 99
[hep-ph/9603244].


\bibitem{uspin} 
R.~Fleischer,
Phys.\ Lett.\ B {\bf 459} (1999) 306;
M.~Gronau,
Phys.\ Lett.\ B {\bf 492} (2000) 297;
T.~Hurth and T.~Mannel,
Phys.\ Lett.\ B {\bf 511} (2001) 196
[hep-ph/0103331] and  hep-ph/0109041.


\bibitem{kly}
C.~S.~Kim, D.~London and T.~Yoshikawa,
Phys.\ Lett.\ B {\bf 458} (1999) 361
[hep-ph/9904311];
A. Datta, C.~S.~Kim and D. London, hep-ph/0105017.


\bibitem{gnw}
Y.~Grossman, Y.~Nir and M.~P.~Worah,
Phys.\ Lett.\ B {\bf 407} (1997) 307
[hep-ph/9704287].

\bibitem{gro-qui}
Y.~Grossman and H.~R.~Quinn,
Phys.\ Rev.\ D {\bf 56} (1997) 7259
[hep-ph/9705356].

\bibitem{kly1}C.~S.~Kim, D.~London and T.~Yoshikawa,
Phys.\ Rev.\ D {\bf 57} (1998) 4010
[hep-ph/9708356].

\bibitem{ddf2}
A.~S.~Dighe, I.~Dunietz and R.~Fleischer,
Phys.\ Lett.\ B {\bf 433} (1998) 147
[hep-ph/9804254].

\bibitem{kayser}
B.~Kayser and D.~London,
Phys.\ Rev.\ D {\bf 61} (2000) 116012
[hep-ph/9909560].




\end{thebibliography}
\end{document}